\documentclass[sigconf]{acmart}

\usepackage{xspace}
\usepackage{graphicx}
\usepackage{subcaption}
\usepackage{multirow}
\usepackage{makecell}

\usepackage[ruled,linesnumbered,noend]{algorithm2e}
\usepackage{hyperref}
\usepackage{listings}
\newcommand{\nickname}{Vec-LUT\xspace}
\newcommand{\reponame}{\texttt{vlut.cpp}\xspace}
\newcommand{\codelink}{\url{https://github.com/OpenBitSys/vlut.cpp}\xspace}
\newcommand{\hflink}{\url{https://huggingface.co/collections/XXXXyu/vlutcpp}\xspace}

\newcommand{\revise}[1]{#1} 
\copyrightyear{2026}
\acmYear{2026}
\setcopyright{cc}
\setcctype{by}
\acmConference[MobiSys '26]{The 24th Annual International Conference on Mobile Systems, Applications and Services}{June 21--25, 2026}{Cambridge, United Kingdom}
\acmBooktitle{The 24th Annual International Conference on Mobile Systems, Applications and Services (MobiSys '26), June 21--25, 2026, Cambridge, United Kingdom}
\acmDOI{10.1145/3745756.3809200}
\acmISBN{979-8-4007-2027-7/2026/06}

\begin{document}

\title{\nickname: Vector Table Lookup for Parallel Ultra-Low-Bit LLM Inference on Edge Devices}


\author{Xiangyu Li}
\authornote{Equal contribution. Sorted by surname in alphabetical order.}
\email{lixiangy22@mails.tsinghua.edu.cn}
\affiliation{
  \institution{Institute for AI Industry Research (AIR), Tsinghua University}
  \city{Beijing}
  \country{China}
}

\author{Chengyu Yin}
\authornotemark[1]
\authornote{Work done during internship at AIR, Tsinghua University.}
\email{cipherxzc@outlook.com}
\affiliation{
  \institution{Beijing Jiaotong University}
  \city{Beijing}
  \country{China}
}

\author{Weijun Wang}
\email{wangweijun@air.tsinghua.edu.cn}
\affiliation{
  \institution{Institute for AI Industry Research (AIR), Tsinghua University}
  \city{Beijing}
  \country{China}
}

\author{Jianyu Wei}
\email{noob@mail.ustc.edu.cn}
\affiliation{
  \institution{University of Science and Technology of China}
  \city{Hefei}
  \country{China}
}

\author{Ting Cao}
\authornote{Corresponding author.}
\email{tingcao@mail.tsinghua.edu.cn}
\affiliation{
  \institution{Institute for AI Industry Research (AIR), Tsinghua University}
  \city{Beijing}
  \country{China}
}

\author{Yunxin Liu}
\email{liuyunxin@air.tsinghua.edu.cn}
\affiliation{
  \institution{Institute for AI Industry Research (AIR), Tsinghua University}
  \city{Beijing}
  \country{China}
}

\renewcommand{\shortauthors}{Xiangyu Li et al.}

\begin{abstract}

Large language models (LLMs) are increasingly deployed on edge devices. To meet strict resource constraints, real-world deployment has pushed LLM quantization from 8-bit to 4-bit, 2-bit, and now 1.58-bit. Combined with lookup table (LUT)-based inference, CPUs run these ultra-low-bit LLMs even faster than NPUs, opening new opportunities for ubiquitous on-device intelligence.

However, this paper identifies that LUT-based inference underutilizes memory bandwidth during parallel inference, which is required for prefilling, test-time scaling, and other multi-token scenarios. The root cause is the scalar LUT paradigm, which performs repetitive and non-contiguous memory accesses for each token.

To solve the issue, we propose vector LUT, a new lookup paradigm that constructs a unified LUT across parallel tokens, and performs a single $1 \rightarrow N$ lookup per index. To realize it efficiently, we further introduce (1) Vector LUT-Centric Tensor Layout, and (2) Cache-Aware Streamed Lookup techniques. Evaluations on 5 edge devices across 3 LLMs show that \nickname outperforms state-of-the-art baselines by up to $4.2\times$. Our implementation is integrated into llama.cpp. The code is available at \codelink.

\end{abstract}





\maketitle


\section{Introduction}
\label{sec:intro}

\begin{figure*}[t]
  \centering
  \includegraphics[trim = 0 0 25 355, clip, width=\linewidth]{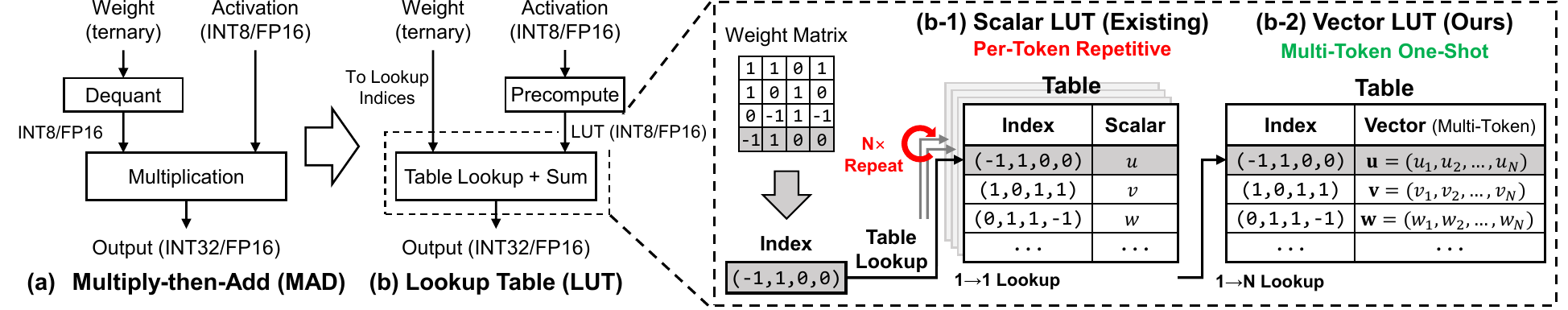}
  \caption{Different mpGeMM paradigms for ternary LLM inference. Our vector LUT stores precomputed results from multiple tokens contiguously in the table, and performs efficient $1\rightarrow N$ lookup, instead of $N\times$ repetitive $1\rightarrow 1$ lookup in existing scalar LUT. Fig.~\ref{fig:lut-example} and \S\ref{sec:bg-lut-preliminary} further explain the mechanism of LUT.}
  \label{fig:loop-order}
\end{figure*}

\begin{figure}[t]
  \centering
  \includegraphics[trim = 0 0 435 245, clip, width=\linewidth]{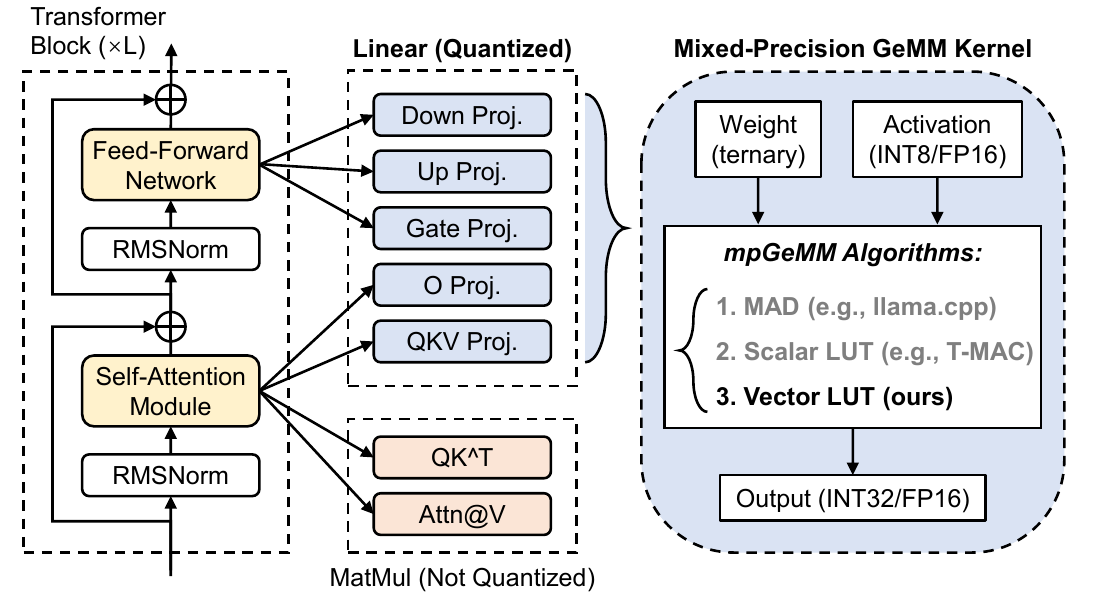}
  \caption{\revise{How mpGeMM kernel optimization integrates into the high-level Transformer inference flow.
      In a quantized LLM, the linear operators, which dominate inference latency in many scenarios, perform mixed-precision multiplication between weights (e.g., ternary) and activations (e.g., INT8/FP16).
      Our method optimizes memory access within the mpGeMM kernel, thereby improving overall inference performance and efficiency.
    }
  }
  \label{fig:flow}
\end{figure}

Large language models (LLMs) are increasingly deployed on edge devices, such as Apple Intelligence on iPhone and Mac~\cite{apple2025intelligence}, Google Gemma 3 on Android~\cite{ikonomidis2025gemma3}, and Microsoft Phi on Copilot PC~\cite{Mehdi2024,kinfey2024phi3intel}. To fit the limited memory budgets of edge devices, quantization has become a de facto technique for on-device LLMs. Recent advances continuously push bit-width from 8-bit~\cite{dettmers2022gpt3,xiao2023smoothquant} to 4-bit~\cite{frantar2022gptq,lin2023awq}, 2-bit~\cite{du2024bitdistiller,chen2025efficientqat} and now 1.58-bit, represented by Microsoft's ternary model and its successors~\cite{ma2024158bitllm,ma2025bitnet,1bitllm2024bitnet,hf1bitllm2024llama3,Falcon3,kaushal2024spectra}. These low-bit models not only demonstrate superior bit-level accuracy scaling~\cite{huang2025tenet,liu2025paretoq}, but also achieve linear speedup via \textit{lookup table (LUT)-based} inference kernels, enabling CPUs (including Apple M2 and Raspberry Pi) to outperform NPUs~\cite{wei2024tmac,wang2025bitnet}. This overturns the need for costly AI accelerators and opens new opportunities for ubiquitous LLM deployment.

\begin{figure}[t]
  \centering
  \includegraphics[trim = 0 0 430 310, clip, width=\linewidth]{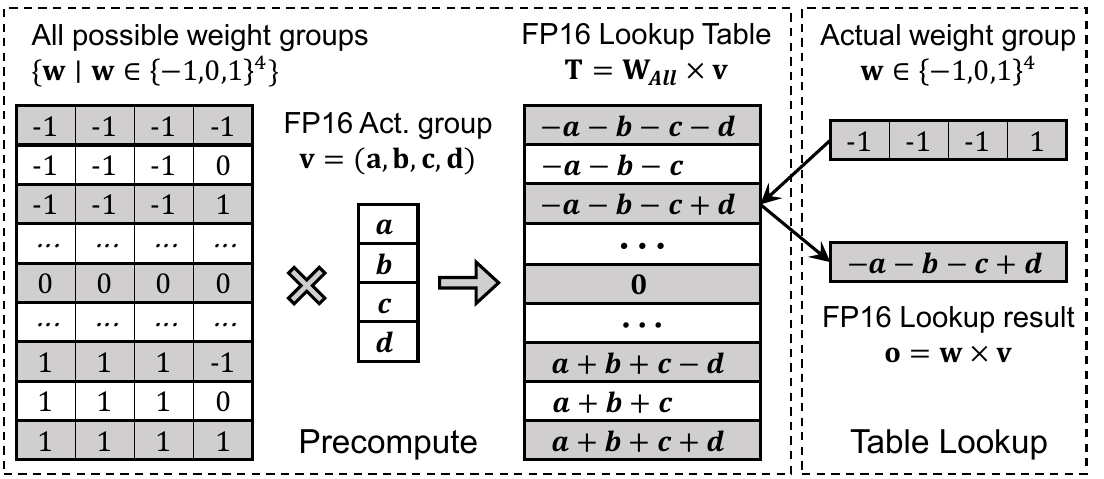}
  \caption{A minimal example of using LUT to calculate $\mathbf{o}=\mathbf{w}\times \mathbf{v}$ with FP16 $\mathbf{v}$ and ternary $\mathbf{w}$ of size 4.}
  \label{fig:lut-example}
\end{figure}

LUT-based inference~\cite{park2022lut,wei2024tmac} was introduced to bridge the gap between low-bit models and hardware without native low-bit support~\cite{ARM2025NEON, Intel2022AVX2,NVIDIA2020AMPERE,Qualcomm2024NPU} (see Fig.~\ref{fig:loop-order}(b), \revise{Fig.~\ref{fig:flow}}, and Fig.~\ref{fig:lut-example}).
The core idea is to partition the weight matrix into small groups, for which all possible bit patterns of a group can be enumerated.
For each bit pattern, its dot product with the activation is precomputed and stored in a lookup table.
This turns matrix multiplication into table lookup indexed by the corresponding bit patterns,
followed by accumulation across groups.
For example, a group of four elements in a ternary model (each weight element ranges \{-1, 0, 1\}) has 81 $(3^4)$ possible bit patterns (e.g., (1, -1, 0, 1) and (-1, 0, 1, 0)), resulting in 81 table entries.

Although LUT-based LLM inference achieves superior performance for single-token generation, we find that it behaves poorly under \textit{parallel inference} (i.e., generating multiple tokens simultaneously), required in scenarios such as prefilling, serving, test-time scaling for decoding~\cite{chen2025parallel,ehrlich2025codemonkeys}, and speculative decoding~\cite{leviathan2023fast,chen2023accelerating}.
In these settings, the memory bandwidth remains highly underutilized, reaching only $\le$ 40\% of its capacity.
This limitation directly harms edge applications that rely on efficient parallel inference, such as multi-app services~\cite{shen2025edgelora,li2025mobilora,yuan2024mobile}, long-context processing agents~\cite{wen2024autodroid,shen2025autoiot}, and streaming video understanding~\cite{ding2025streammind}.

By analyzing this behavior, we find that the inefficiency stems from the \textit{$1\rightarrow1$ table lookup} paradigm, which we refer to as \textit{scalar LUT} (see Fig.~\ref{fig:loop-order}(b-1)).
During parallel inference, each token’s activation precomputes its own tables. Hardware LUT instructions (e.g., the ARM TBL instruction) are then called to perform table lookup. An $N$-token parallel inference requires loading $N$ tables and executing lookup for $N$ times. Since memory access of table lookup is inherently random, and the table size (e.g., hundreds of MiBs) far exceeds cache capacity, this repetitive non-contiguous memory access leads to severe memory-bandwidth underutilization and substantially increased latency.

To address this issue, we propose a novel \textit{vector LUT paradigm that performs a single $1\rightarrow N$ lookup for each index, rather than performing $1\rightarrow 1$ lookup for $N$ times}. Instead of constructing a separate table for each token, vector LUT precomputes a unified table, as shown in Fig.~\ref{fig:loop-order}(b-2), where the results corresponding to the same index are stored contiguously across tokens. It eliminates reliance on hardware scalar LUT instructions, avoids non-contiguous memory access, and potentially improves memory bandwidth utilization.

While this idea is conceptually clean, realizing it effectively requires overcoming two technical challenges.

\paragraph{({\romannumeral 1}) How to design tensor layouts for memory bandwidth efficiency}
Since vector LUT constructs a unified table across all tokens rather than separated ones for each token, all tensors involved in the computation, including weight, activation, and LUT, should be organized to match this paradigm for contiguous memory access. In scalar LUT, each token maintains its own table, leading to an activation and weight layout that is feature-first.
The unified table used by vector LUT requires a token-first layout so that all tokens’ values for the same feature are stored contiguously.
If tensors are misaligned with this requirement, non-contiguous memory accesses can degrade performance by up to $12\times$ (see~\S\ref{sec:eval-ablation}).

\paragraph{({\romannumeral 2}) How to design the LUT access to avoid cache thrashing}
As shown in Fig.~\ref{fig:loop-order}, vector LUT stores the results for $N$ tokens in each row, which expands the LUT size by a factor of $N$, making it infeasible to cache the entire LUT for fast lookup.
For example, while typical L1 and L2/L3 caches for edge processors are only 10s of KiBs and 10s of MiBs, respectively~\cite{arm_neoverse_v1, intel_core_i7_13700k}, a 2-bit vector LUT in INT16 can exceed $200$ MiB
(e.g., $285.5$ MiB for a $4096\times 14436\times 512$ GeMM from Llama3 8B).
During lookup, the large LUT combined with the inherently random access pattern can lead to a poor cache hit rate and low inference throughput.

To address the challenges, we present \nickname, an efficient mpGeMM kernel design (mixed-precision GeMM, e.g., W2A16 and W1.6A8), built upon the vector LUT paradigm for parallel inference. Technically, we propose:
({\romannumeral 1}) \textit{Vector LUT-Centric Tensor Layout}, which reorganizes weight, activation, and LUT tensors into a token-contiguous rather than feature-contiguous layout. The online reorganization is fused into LUT operation, with minimal overhead while compatible with existing inference frameworks.
Since vector LUT eliminates the need for lookup instructions, we can pack weights as flexible decimal indices.
({\romannumeral 2}) \textit{Cache-Aware Streamed Lookup}, which prevents cache thrashing by performing table lookup within cache-friendly tiles. This design streamlines the precompute-lookup process: instead of generating and storing the entire LUT beforehand, each tile is precomputed on-demand and then looked up entirely within cache.

We implement and integrate \nickname into llama.cpp~\cite{llamacpp}, with particular support for SOTA accuracy-bit-efficient ternary models. Our lossless weight packing can achieve b1.60\footnote{The number indicates the BPW (bits per weight)} precision (\texttt{I1}) for ternary models, beyond b2.00 (\texttt{I2}), providing the most compact ternary packing to date.
We conduct a comprehensive evaluation on 5 edge devices and 3 LLMs (BitNet~\cite{1bitllm2024bitnet}, Llama 3~\cite{hf1bitllm2024llama3}, and Falcon~\cite{Falcon3}).
Compared to SOTA frameworks (T-MAC~\cite{wei2024tmac}, bitnet.cpp~\cite{ma2025bitnet}, and llama.cpp~\cite{llamacpp}) requiring only CPUs, \nickname achieves up to $4.2\times$ (\texttt{I1}) and $2.6\times$ (\texttt{I2}) speedup\revise{, saving energy by up to 2.1$\times$}.
\revise{
  Moreover, \nickname achieves a throughput of 273.5 tokens/s on an affordable CPU server (\$0.50/h on AWS) under continuous batching workloads.
  Using only 2 CPU cores on Snapdragon 8 Elite, \nickname even outperforms the NPU-based solution (llama.cpp's Hexagon backend) by 1.1$\times$, despite the device's powerful NPU.
}

We summarize our main contributions as follows:

\scalebox{0.8}{$\bullet$} We propose vector LUT, a new paradigm for LUT-based inference. It constructs a unified LUT across all parallel tokens and performs a single $1 \rightarrow N$ lookup per weight index.

\scalebox{0.8}{$\bullet$} We present \nickname, an efficient mpGeMM kernel built on the vector LUT paradigm, optimized with Vector LUT-Centric Tensor Layout and Cache-Aware Streamed Lookup.

\scalebox{0.8}{$\bullet$} We implement \nickname with llama.cpp integration, outperforming SOTA LUT baselines by up to $4.2\times$.
\section{Background and Motivation}

\begin{figure}[t]
  \centering
  \includegraphics[width=0.9\linewidth]{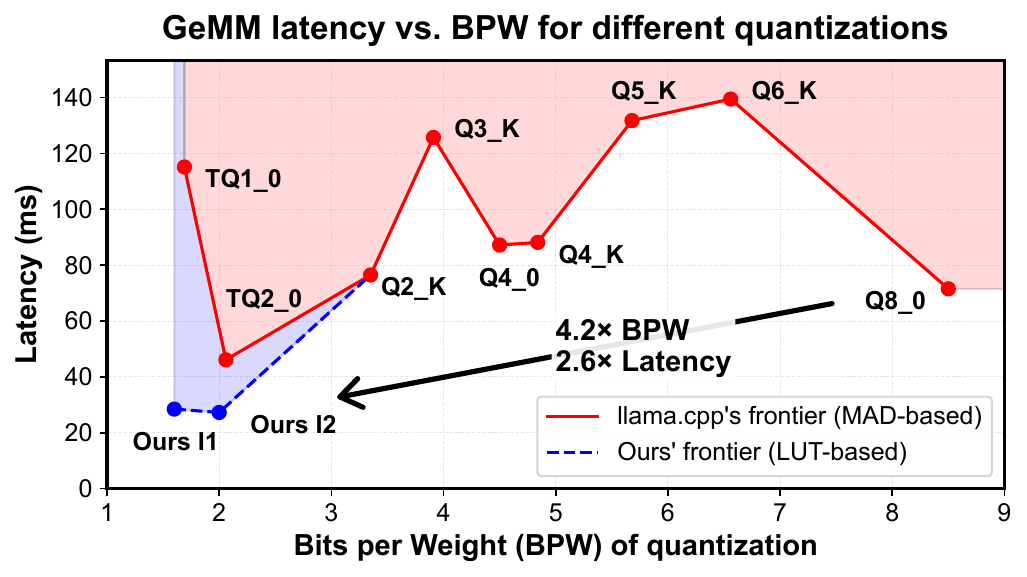}
  \caption{mpGeMM latency vs. BPW on an Intel PC. \nickname utilizes fewer BPWs (i.e., $\le2$) for lower latency, while MAD-based llama.cpp cannot achieve speedup with fewer BPWs.}
  \label{fig:gemm-bits}
\end{figure}

\subsection{Parallel Inference and Ultra-Low-Bit LLMs}

\label{sec:bg-parallel}

\subsubsection{Edge applications of parallel LLM inference}

\textit{Parallel inference}, which processes multiple tokens in parallel, is one of the most critical workloads in today's LLM-based applications.
It is predominantly exercised in two scenarios: prefilling and parallel decoding.
\textbf{({\romannumeral 1}) Prefilling} of LLMs ingests multiple tokens (e.g., in a user prompt) simultaneously as input.
Its latency directly affects the responsiveness of edge applications with long context inputs, such as GUI agents~\cite{wen2024autodroid} and AIoT agents~\cite{shen2025autoiot}.
\textbf{({\romannumeral 2}) Parallel decoding} of LLMs generates multiple tokens simultaneously.
Its performance not only determines the overall QoS in multi-app services~\cite{shen2025edgelora,li2025mobilora}, but also influences the effective decoding latency in parallel test-time scaling~\cite{ehrlich2025codemonkeys, chen2025parallel} and speculative decoding~\cite{leviathan2023fast, chen2023accelerating}.
\revise{Accelerating parallel inference on edge devices is therefore essential to meet the latency and throughput requirements of these applications, enabling the deployment of more capable models under resource constraints.}

\subsubsection{Superiority of ternary LLMs in edge scenarios}

As the bit-width of LLMs continues to shrink, ternary-valued models, where each weight takes value from \{1,-1, 0\}, emerge as a Pareto-optimal point in the accuracy-cost tradeoff. Their advantages manifest in two dimensions.
\textbf{({\romannumeral 1}) Theoretical superiority.} Recent research demonstrates the superior {\itshape accuracy per joule}~\cite{huang2025tenet} and {\itshape accuracy per bit trade-offs}~\cite{liu2025paretoq} of ternary quantization. For example, Microsoft's BitNet 2B costs only 0.4 GB memory~\cite{ma2025bitnet}, but 4-bit models of similar accuracy require $\geq$ 1 GB.  Also, TENET~\cite{huang2025tenet} and LUT Tensor Core~\cite{mo2024lut} demonstrate the superior latency/energy efficiency of ternary LLMs with co-designed hardware.
\textbf{({\romannumeral 2}) Practical benefits.} Past deployments from Microsoft (e.g., BitNet.cpp and T-MAC) have demonstrated ternary models can run smoothly on diverse x86/ARM/iOS edge devices using CPUs alone, including mobile phones, desktops (Microsoft Surface Book, Apple M2) and even Raspberry Pi. The running speed can even surpass NPUs. For example, T-MAC reports that a 4-bit 7B model reaches 18.7 tokens/s on Snapdragon X Elite CPUs, whereas the NPU attains only 10.4 tokens/s~\cite{wei2024tmac}.
This effectively removes the dependence on GPUs/NPUs for LLM inference and opens significant potential for deployment across edge and IoT devices.

\subsubsection{LUT-based Ultra-Low-Bit LLM inference}

\label{sec:bg-lut-preliminary}

LUT-based inference replaces runtime dequantization and multiplication with table lookup.
As shown in Fig.~\ref{fig:lut-example}, take a ternary model with weights partitioned in groups of 4 as an example, it enumerates all possible bit patterns of weight groups (i.e., $\mathbf{w} \in \{-1, 0, 1\}^4$), and then precomputes their dot products with the activation group $\mathbf{v}$.
The precomputed results are stored in the table $\mathbf{T}=\{\mathbf{w} \times \mathbf{v} \mid \mathbf{w} \in \{-1, 0, 1\}^4 \}$, obtaining $3^4=81$ table entries in total.

Then it traverses the actual weight matrix, and uses the weight groups as indices to directly look up the dot product $\mathbf{o}=\mathbf{w} \times \mathbf{v}=\mathbf{T}(idx(\mathbf{w}))$ without multiplication.
LUT is suitable for low-bit, especially ternary models, for which the possible bit patterns in $\mathbf{w}$ are limited to enumerate (e.g., $3^4=81$ for groups of 4).
On top of this basic LUT design, our vector LUT paradigm constructs the table in the granularity of vectors, so each table entry stores precomputed results from multiple tokens contiguously.

\subsection{Limitations of Existing Approaches}

\label{sec:bg-mpgemm}

We conduct in-depth profiling of both MAD-based (Multiply-Add) and scalar LUT-based mpGeMM kernels to diagnose their bottlenecks, and motivate our vector LUT design.

\subsubsection{Costly Dequantization in MAD-based mpGeMM}

\label{sec:bg-mad-limitation}

Conventional MAD-based mpGeMM must dequantize low-bit weights to hardware-supported precisions and then multiply them with activations in aligned precisions (e.g., INT8/FP16).
The dequantization and multiplication overheads counteract the benefit of low-bit quantization, and even increase the latency, as demonstrated in Fig.~\ref{fig:gemm-bits}\footnote{Q\# in the figure denotes the quantization formats used in llama.cpp, where \# indicates the bits per weight (BPW). TQ\# refers to the corresponding ternary quantization formats in llama.cpp.}.
Among MAD-based mpGeMM kernels in llama.cpp, the higher-BPW 8-bit kernel (\texttt{Q8\_0}, \revise{dequantization-free}) delivers lower latency than the 4-bit kernel (\texttt{Q4\_0}), and the 2-bit (\texttt{TQ2\_0}) ternary packing outperforms the sub-2-bit (\texttt{TQ1\_0}) one significantly.
In contrast, our LUT-based approach directly leverages compactly packed weights for table lookup without dequantization or unpacking (more in \S\ref{sec:design-layout}), yielding remarkable acceleration at low BPWs.

\begin{table}\small
  \caption{Time breakdown of T-MAC's mpGeMM kernel. Measured on an Orange Pi 5 Plus \revise{with a single thread}.}
  \label{tab:tmac-breakdown}
  \begin{tabular}{ccccc}
    \toprule
    $M\times K$ & Precompute & \textbf{Lookup} & Accumulate & Scale \\
    \midrule
    $320\times 3200$ & 0.8\% & \underline{\textbf{47.6\%}} & 25.0\% & 26.6\% \\
    $128\times 8640$ & 0.7\% & \underline{\textbf{47.3\%}} & 25.5\% & 26.4\% \\
    \bottomrule
  \end{tabular}
\end{table}


\subsubsection{Repetitive Lookup in Scalar LUT}

\label{sec:bg-lut-limit}

Existing LUT-based kernels (e.g., T-MAC) can fully leverage the memory bandwidth for single-token generation. However, we observe a severe bandwidth underutilization (<40\%) for parallel inference. We identify the reason is that existing LUT-based kernels adopt the scalar LUT paradigm ($1\rightarrow1$ table lookup), which constructs an independent table for each token, and then calls hardware lookup instructions (e.g., SIMD TBL instruction on ARM CPU and PSHUF on x86 CPU) to look up the table.\revise{We find that this paradigm fundamentally cannot achieve efficient reuse of both tables and indices simultaneously.}

\revise{
  During parallel inference with $N$ tokens, the kernel must construct $N$ tables, each randomly accessed $\mathcal{O}(MK)$ times.
  T-MAC reuses each table across all indices before loading the next, but incurs $N\times$ repetitive table loading and lookup with non-contiguous memory access within tables.
  The alternative---reusing each index across tables---would cause non-contiguous memory access across tables or even frequent LUT swapping, with even higher overhead.
  Both reuse strategies in scalar LUT lead to severe memory-bandwidth underutilization and substantially increased latency.
}
In some cases, scalar-LUT kernels even underperform MAD-based kernels (more in \S\ref{sec:eval-gemm}).

As shown in Table~\ref{tab:tmac-breakdown}, even optimized with hardware instructions, ``Lookup'' (including weight loading and table lookup) still takes up nearly half of T-MAC's \revise{single-thread} mpGeMM latency.
In contrast, \nickname will reduce the lookup cost to below 1\% (details in \S\ref{sec:eval-breakdown}).

\subsection{Insight and Opportunities}
\label{sec:bg-cause-insight-opportunity}

\subsubsection{Unified table shared by all tokens}
\label{sec:bg-insight}

Based on the analysis in \S\ref{sec:bg-lut-limit}, we propose the vector LUT paradigm.
Our key insight is that instead of constructing a separate table for each token, we build a unified table shared across all $N$ parallel tokens, \textbf{replacing $N$ independent $1 \rightarrow 1$ lookups with a single $1 \rightarrow N$ lookup}. This transformation eliminates repeated non-contiguous memory accesses, converts them into contiguous accesses for each weight index, and removes the dependency on hardware LUT instructions.
Given the high proportion of “Lookup’’ cost in T-MAC (Table~\ref{tab:tmac-breakdown}), vector LUT is expected to deliver up to $2\times$ kernel-level speedup by effectively eliminating lookup overhead. \nickname reduces lookup cost to below 1\% in \S\ref{sec:eval-breakdown}.

\subsubsection{Other opportunities by not relying on hardware lookup instructions}
\label{bg-other-opportunity}

Besides the key opportunity in \S\ref{sec:bg-insight}, vector LUT brings two more potential gains by not relying on hardware lookup instructions.
\textbf{({\romannumeral 1}) Accumulation reduction with larger indexing range.}
To satisfy the bit-width requirements of SIMD hardware lookup instructions, existing kernels only allow a $2^4$ indexing range with a 4-bit index.
In contrast, vector LUT allows up to a $3^5$ indexing range with an 8-bit index (details in Fig.~\ref{fig:bit-map}).
It reduces the required lookup and accumulation count by $\sim2\times$ (twice lookup and accumulation for 4-bit index compared to 8-bit index), providing up to $1.25\times$ additional kernel speedup (see Table~\ref{tab:tmac-breakdown} accumulate cost).
\textbf{({\romannumeral 2}) Enhanced versatility and efficiency with flexible sub-2-bit packing.}
To utilize SIMD acceleration,
existing MAD-based kernels impose strict weight shape requirements (e.g., by multiples of 256 in llama.cpp's \texttt{TQ1\_0} and \texttt{TQ2\_0}~\cite{compilade2024tq}), which significantly limits their generality.
In contrast, vector LUT can pack weights of almost {\itshape any shapes} more compactly for memory saving, without compromising performance or accuracy.

\section{Design}

\label{sec:design}

Based on the insight in \S\ref{sec:bg-cause-insight-opportunity}, we propose the vector LUT paradigm for parallel ultra-low-bit LLM inference, and design the \nickname mpGeMM kernel (\S\ref{sec:design-overview} and \S\ref{sec:design-algorithm}). 

To effectively support this paradigm, we also need to solve the challenges of ({\romannumeral 1}) tensor layout design to align with the vector LUT paradigm, and ({\romannumeral 2}) the LUT access design to avoid cache thrashing. We propose two optimizations accordingly: ({\romannumeral 1}) {\itshape Vector LUT-Centric Tensor Layout} that improves memory bandwidth utilization with a contiguous layout (details in \S\ref{sec:design-layout}, and ({\romannumeral 2}) {\itshape Cache-Aware Streamed Lookup} that enhances cache hit and register reuse with cache-aware tiling and streamed precomputing-lookup execution (details in \S\ref{sec:design-stream}).




\begin{figure}[t]
    \centering
    \includegraphics[trim = 0 0 560 65, clip, width=0.8\linewidth]{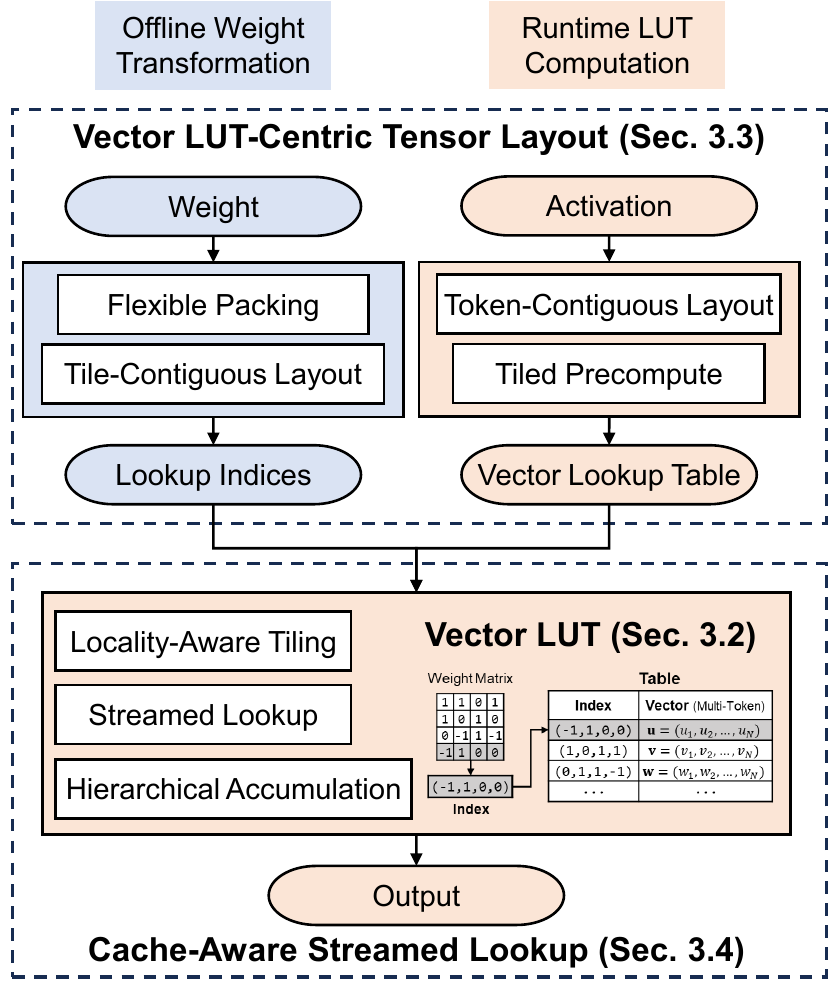}
    \caption{Overview of the \nickname mpGeMM kernel.}
    \label{fig:overview}
\end{figure}

\subsection{System Overview}

\label{sec:design-overview}

Figure~\ref{fig:overview} shows the overview of \nickname's mpGeMM kernel.
It involves two stages: {\itshape offline weight transformation} and {\itshape runtime LUT computation}.
\textbf{({\romannumeral 1}) At the offline stage}, \nickname flexibly packs floating-point weights to low-bit integers, and then permutes them to a tile-contiguous layout. The transformed weights will be directly used as lookup indices at runtime.
\textbf{({\romannumeral 2}) At the runtime stage}, \nickname transposes feature-contiguous activations (common in inference frameworks like llama.cpp) to token-contiguous, and then performs LUT precomputing and lookup in tiles.
Specifically, the vector LUT contains precomputed results from the activation (introduced in \S\ref{sec:bg-lut-preliminary}), and each lookup index (transformed from weights) maps to a vector of precomputed results (i.e., $1\rightarrow N$ lookup).

Our design includes the core algorithm (\S\ref{sec:design-algorithm}), layout optimizations (\S\ref{sec:design-layout}), and lookup scheme optimizations (\S\ref{sec:design-stream}).
The offline stage mainly involves layout optimizations, including flexible sub-2-bit packing and tile-contiguous packed weight layout, which will be introduced in \S\ref{sec:design-layout}.
The runtime stage combines layout and lookup scheme optimizations.
Layout optimizations, including token-contiguous LUT layout and the corresponding activation and output transformation, will be introduced in \S\ref{sec:design-layout}.
Lookup scheme optimizations, including locality-aware LUT tiling, streamed precomputing\allowbreak-lookup execution and INT16-INT32 hierarchical accumulation, will be introduced in \S\ref{sec:design-stream}.

\begin{table}[t]\small
  \centering
  \caption{Notations for mpGeMM: $\mathbf{W} \times \mathbf{A} = \mathbf{O}$.}
  \label{tab:gemm-notation}
  \setlength{\tabcolsep}{4pt}
  \begin{tabular}{cl}
    \toprule
    \textbf{Tensor} & \textbf{Description} \\
    \midrule
    $\mathbf{W} \in \mathbb{R}^{M \times K}$ & Weight \\
    $\mathbf{A} \in \mathbb{R}^{K \times N}$ & Activation \\
    $\mathbf{O} \in \mathbb{R}^{M \times N}$ & Output \\
    \bottomrule
  \end{tabular}
  \quad
  \vline
  \quad
  \begin{tabular}{cl}
    \toprule
    \textbf{Dimension} & \textbf{Description} \\
    \midrule
    $M$ & Out. Feature \\
    $K$ & In. Feature \\
    $N$ & Token Length \\
    \bottomrule
  \end{tabular}
\end{table}

\subsection{Vector LUT Algorithm}

\label{sec:design-algorithm}

\begin{algorithm}[t]\small
  \caption{\nickname's Core Algorithm}
  \label{alg:lut}

  
  \SetKwInOut{Input}{input}
  \SetKwInOut{Output}{output}

  \Input{Activation $\mathbf{A}$ of shape $K\times N$,\\
         Packed weights $\mathbf{W}$ of shape $M\times K/g$\\
         where $g$ is the group size of packed weights.}
  \Output{Result $\mathbf{O}$ of shape $M\times N$}

  \BlankLine
  $\mathbf{T} \leftarrow \text{Precompute}(\mathbf{A}, K, N)$\;

  \BlankLine
  \For{$k \leftarrow 1$ \KwTo $K/g$}{
    \For{$m \leftarrow 1$ \KwTo $M$}{
      $idx \leftarrow \mathbf{W}[m,k]$\tcp*{Shared index among $N$}
      \For{$n \leftarrow 1$ \KwTo $N$}{
        $\mathbf{O}[m,n] \text{ += } \mathbf{T}[k,idx,n]$\tcp*{Vector add}
      }
    }
  }

  \BlankLine
  \SetKwFunction{FPrecompute}{Precompute}
  \SetKwProg{Fn}{Function}{:}{}
  \Fn{\FPrecompute{$\mathbf{A}, K, N$}}{
    \For{$k \leftarrow 1$ \KwTo $K/g$}{\tcp*{Sub-tables}
      \For{$i, j \leftarrow 1$ \KwTo $3^g, g$}{
        $v \leftarrow \text{GetSign}(i,j)$\tcp*{Enumerated index}
        \If{$v = 1$}{
          \For{$n \leftarrow 1$ \KwTo $N$}{
            $\mathbf{T}[k, i, n] \text{ += } \mathbf{A}[kg+j,n]$\tcp*{Add}
          }
        }
        \ElseIf{$v = -1$}{
          \For{$n \leftarrow 1$ \KwTo $N$}{
            $\mathbf{T}[k, i, n] \text{ -= } \mathbf{A}[kg+j,n]$\tcp*{Sub}
          }
        }
        \Else{
          Continue\tcp*{Nop}
        }
      }
    }
    \Return{LUT}
  }

  \BlankLine
  \SetKwFunction{FGetSign}{GetSign}
  \SetKwProg{Fn}{Function}{:}{}
  \Fn{\FGetSign{$i, j$}}{
    $trit \leftarrow \lfloor i / 3^{j-1} \rfloor \bmod 3$\;
    \Return{$trit-1$}\tcp*[f]{$j$-th trit (in $\{-1,0,1\}$) of $i$}
  }
\end{algorithm}

\begin{figure*}
  \centering
  \includegraphics[trim = 0 0 20 200, clip, width=0.8\linewidth]{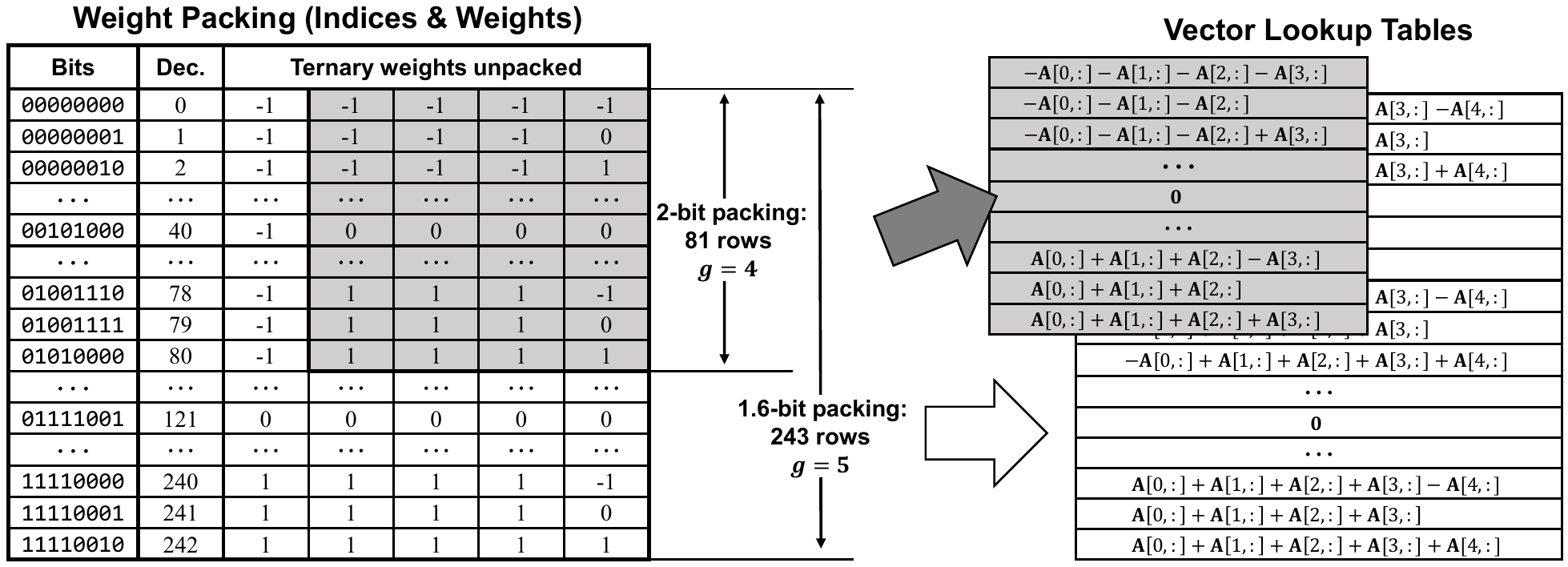}
  \caption{Mappings among packed weights (bits and decimal), unpacked weights (ternary), and precomputed LUT rows. The LUT row index is equal to the decimal value of packed weights, avoiding extra conversion.}
  \label{fig:bit-map}
\end{figure*}

The Vector LUT-based mpGeMM algorithm (described in Alg.~\ref{alg:lut}) is the core of \nickname.
It includes two steps: {\itshape LUT precompute} and {\itshape Table lookup \& Accumulate}. We'll introduce the details below.

\textbf{\itshape LUT precompute} ingests the activation and builds the vector LUT.
For an activation $\mathbf{A}$, of shape $K\times N$, \nickname traverses the dimension $K$ of $\mathbf{A}$ in groups of $g$ (line 8), and precomputes a \textit{sub-table} of shape $3^g\times N$ for each group (line 9 to 18).
This produces a unified table $\mathbf{T}$ containing $K/g$ sub-tables.
In $\mathbf{T}$ of shape $K/g\times 3^g\times N$, $\mathbf{T}[k,i,n]$ is calculated by:
\begin{equation}
  \mathbf{T}[k,i,n]=\sum_{i=1}^{3^g}\sum_{j=1}^{g} \text{GetSign}(i,j)\mathbf{A}[kg+j, n],
\end{equation}
where, GetSign$(i,j)$ extracts the $j$-th trit (i.e., ternary value, ranging $\{-1,0,1\}$) from an integer $i$, which represents the operation (i.e., subtraction (sub), nop, or addition (add)) to the $j$-th row in the $k$-th group of activation $\mathbf{A}$, to calculate the $i$-th row of results in the $k$-th sub-table of $\mathbf{T}$.
For example, when $g=4$, $i = 23 = 0\times3^3 + 2\times3^2 + 1\times3 + 2\times1 \rightarrow (0,2,1,2) \rightarrow (-1, 1, 0, 1) \rightarrow (\text{sub, add, nop, add})$.
Fig.~\ref{fig:bit-map} shows a complete view of mappings between indices and corresponding ternary weights, for both $g=4$ (2-bit packing) and $g=5$ (1.6-bit packing).
As a result, each sub-table in $\mathbf{T}$ contains $3^g$ table entries of length $N$, representing the $3^g$ possible combinations (with sub, nop, and add) of the $g$ rows in the $k$-th group of activation $\mathbf{A}$.

After precomputing $\mathbf{T}$, \textbf{\itshape Table lookup \& Accumulate} traverses the transformed weights $\mathbf{W}$ in tiles, and uses each packed byte as an index for table lookup.
Fig.~\ref{fig:bit-map} shows the mappings from indices to results (i.e., corresponding rows in the vector LUT).
Continuing with the $i = 23$ example, \nickname looks up the table with index 23 and gets $-\mathbf{A}[0,:]+\mathbf{A}[1,:]+\mathbf{A}[3,:]$, which is the precomputed result for this specific weight pattern ($\{-1, 1, 0, 1\}$).
The results are then accumulated to obtain the token-contiguous output $\mathbf{O}$ of shape $M\times N$.
Specifically, $\mathbf{O}[m,n]$ is calculated by:
\begin{equation}
  \mathbf{O}[m,n] = \sum_{k=1}^{K/g} \mathbf{T}[k,\mathbf{W}[m,k],n], 
\end{equation}
in which, each packed index (of \texttt{uint8}) in $\mathbf{W}$ ranges from 1 to $3^g$, where $g\in \{4,5\}$~\footnote{To be precise, the \texttt{uint8} index range is $[0,3^g)$ (Dec. in Fig.~\ref{fig:bit-map}).}.

The core difference between vector LUT and existing scalar LUT is that \nickname precomputes (line 4 to 6) and looks up (line 10 to 18) the LUT in the granularity of vectors, instead of single elements.
This significantly reduces random non-contiguous memory access in scalar LUT, with the technical challenges solved by \S\ref{sec:design-layout} and \S\ref{sec:design-stream}.

\subsection{Vector LUT-Centric Tensor Layout}

\label{sec:design-layout}

\nickname adopts the {\itshape Vector LUT-Centric Tensor Layout} to keep memory access contiguous in vector LUT.
It includes token-contiguous layout for activations, LUTs, and outputs, and tile-contiguous layout for weights.
Besides, we propose flexible sub-2-bit packing for a compact weight layout without shape restriction and performance degradation.

\paragraph{Token-contiguous LUT layout}
To improve the memory bandwidth utilization during lookup and accumulation, \nickname layouts the LUT $\mathbf{T}$ of shape $K/g\times 3^g\times N$ in row-major, keeping the token dimension $N$ contiguous.
This ensures contiguous memory access during vector addition in the inner loop (e.g., line 6 in Alg.~\ref{alg:lut}),
and leads to an up to $12\times$ speedup, when applied with the weight layout below (more in \S\ref{sec:eval-ablation}).
The activation $\mathbf{A}$ and accumulated output $\mathbf{O}$ also adopt this token-contiguous layout to keep aligned with the LUT.

\paragraph{Tile-contiguous packed weight layout}
To further reduce the accumulation count and improve memory bandwidth utilization, \nickname adopts the tile-contiguous packed weight layout.
It packs the ternary weights into bytes (e.g., \texttt{uint8}) for direct LUT indexing (details in Fig.~\ref{fig:bit-map}).
This byte-wise packing bypasses the bit-width limitation in existing kernels, leading to a remarkable reduction of the accumulation count, as mentioned in \S\ref{sec:bg-cause-insight-opportunity}.
Then, each weight tile (details in \S\ref{sec:design-stream}) is flattened and contiguously stored in bytes to ensure tile-wise contiguous access.
The tiles are then permuted by the input feature ($K$) to keep aligned with the $K\rightarrow M$ loop order in Alg.~\ref{alg:lut}.

\paragraph{Flexible sub-2-bit weight packing}
Unlike existing kernels which have strict shape requirements (mentioned in \S\ref{bg-other-opportunity}), \nickname adopts a flexible and lossless sub-2-bit packing method to further reduce the memory footprint without performance degradation.
Fig. \ref{fig:bit-map} shows \nickname's two basic packing methods, the 1.6-bit ($g=5$) and 2-bit ($g=4$) packing, which already have much looser shape requirements (i.e., $g \mid K$) compared to existing kernels (e.g., llama.cpp).
\nickname further combines these two packing methods by applying different packings for different groups of weights.
This further expands the weight shape support to almost any $K$ values~\footnote{The condition is $(4a+5b)\mid K, a,b \in \mathbb{Z}, ab>0$.}, and always achieves a near-1.6-bit BPW for common weight shapes.

\paragraph{Fused activation and output transformation}
In most inference frameworks (e.g., llama.cpp), the activations and outputs are in a feature-first layout, which is different from \nickname's token-first layout.
Therefore, to integrate into these frameworks, \nickname needs extra transposition and reverse-transposition for activations and outputs, respectively.
To minimize the overhead of these extra operations, we fuse them into LUT precomputing and results accumulation, achieving significant end-to-end inference acceleration (up to $4.2\times$ in prefilling and $3.2\times$ in parallel decoding in \S\ref{sec:eval-e2e}).

\subsection{Cache-Aware Streamed Lookup}

\label{sec:design-stream}

\begin{figure}
  \centering
  \includegraphics[trim = 0 0 570 370, clip, width=\linewidth]{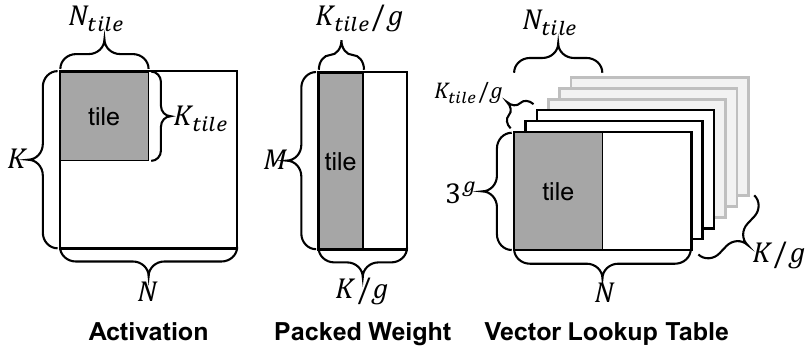}
  \caption{Tiled tensors in \nickname. The LUT is tiled by sub-tables ($K/g$) and tokens ($N$) for locality.}
  \label{fig:tile}
\end{figure}

\nickname adopts the {\itshape Cache-Aware Streamed Lookup} scheme to address the cache thrashing issue caused by large LUT storage.
It includes locality-aware LUT tiling and the streamed precomputing-lookup execution that perform LUT-related operations in high-speed caches, and INT16-INT32 hierarchical accumulation that improves the accumulation throughput within limited cache bandwidth.

\paragraph{Locality-aware LUT tiling}
To reduce the LUT memory footprint (hundreds of MiBs if not tiled) while reducing random access during lookup, \nickname adopts a locality-aware tiling strategy.
As shown in Fig.~\ref{fig:tile},
it tiles the LUT by sub-tables ($K/g$) with a tile size of $K_{tile}/g$, and by tokens ($N$) with a tile size of $N_{tile}$.
The $3^g$ rows of sub-tables are not tiled as they will be randomly accessed during table lookup.
This reduces the LUT memory footprint by up to $(K/K_{tile})(N/N_{tile})\times$, and the optimal tile size depends on hardware specifications (details in \S\ref{sec:impl}).
In our implementation, it achieves KiB level LUT size per tile, which is sufficient to fit in typical L1 cache, achieving less than 10\% cache-miss overhead in VTune~\cite{Intel2025VTune} profiling.

\paragraph{Streamed precomputing-lookup execution}
To reduce memory copy from main memory to cache, we take a further step to streamline LUT precomputing and lookup by tiles.
Since each tile is sufficiently small to fit in high-speed cache, \nickname can conduct all LUT-related memory access solely in cache.
Compared to precomputing the entire table in the main memory (existing kernels' practice), this streamed execution provides up to $3\times$ inference speedup, combined with the tiling design above (evaluated in \S\ref{sec:eval-ablation}).

\paragraph{INT16-INT32 hierarchical accumulation}
Besides the locality issues, another source of the memory pressure is the precision mismatch during accumulation.
\revise{
Following llama.cpp and the models' official implementations on Huggingface, the activations are per-token symmetrically quantized to INT8 for input, with results accumulated in INT32 for output.
This avoids overflow but reduces arithmetic throughput by $4\times$ compared to INT8.
}
To address this, we adopt a hierarchical accumulation design that uses INT16 to bridge INT8 and INT32:
({\romannumeral 1}) LUT precomputing from INT8 (activations) to INT16 (LUT elements).
({\romannumeral 2}) Intra-block accumulation in INT16 (LUT elements and block-wise intermediates).
\revise{It doesn't involve clipping or scaling, and} won't cause overflow as long as the block size $B\le\lfloor \frac{max(\text{INT16})}{max(\text{INT8}) \times g}\rfloor$.
({\romannumeral 3}) Inter-block accumulation from INT16 (intermediates) to INT32 (outputs).
With sufficiently large $B$ (e.g., $B=64$ when $g=4$), this hierarchy significantly improves accumulation throughput\revise{, without introducing numerical error}.

\section{Implementation}

\label{sec:impl}

We implement and integrate \nickname into llama.cpp, the most popular open-source LLM inference framework on edge devices with SOTA performance.
\nickname supports two lossless packing methods: \texttt{I1} (b1.60) and \texttt{I2} (b2.00).
\revise{We integrate them as two custom quantization types in llama.cpp and implement our optimizations within the mpGeMM kernel, without modifying the high-level inference flow.}
Due to the tensor layout mismatch, we implement the optional activation and output transformation steps, and fuse them with LUT precomputing and accumulation, as described in \S\ref{sec:design-layout}.

\begin{figure}[t]
    \centering
    \includegraphics[trim = 0 0 300 380, clip, width=\linewidth]{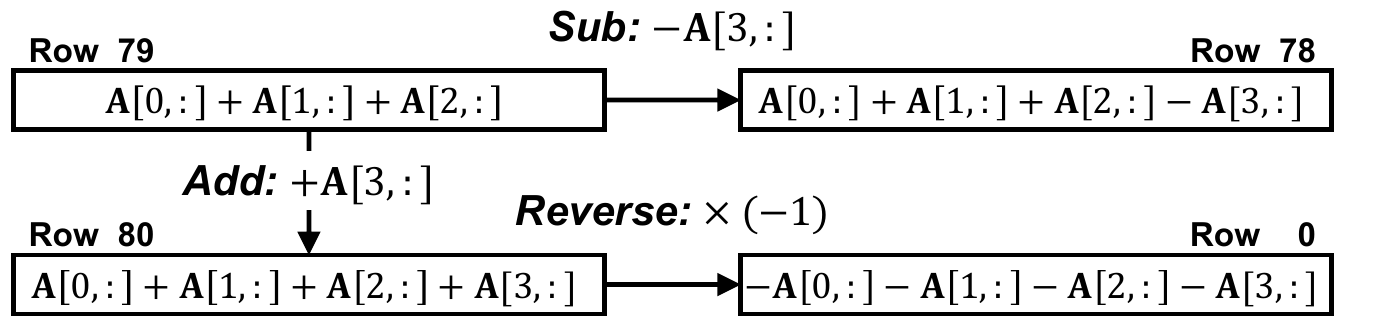}
    \caption{Examples of topological precomputing to reduce redundant calculations by reusing intermediates. Dependencies of LUT entries: $79\rightarrow78, 79\rightarrow80\rightarrow0$.}
    \label{fig:topological-precompute}
\end{figure}

\paragraph{Topological precomputing}
To further reduce the LUT precomputing cost, we adopt a topological precomputing method that minimizes the operations required.
It's based on the observation that there are many repetitive calculations when precomputing different entries in the LUT.
\nickname avoids them by reusing intermediates, as shown in Figure~\ref{fig:topological-precompute}.
Topological precomputing reduces the number of operations from $2\times 3^{g-1}g$ (as in the vanilla Alg.~\ref{alg:lut}) to $3^g$, providing a $(2g/3)\times$ precomputing speedup (e.g., by $3.3\times$ when $g=5$).

\paragraph{Tile size selection}
\revise{\nickname determines all system parameters offline, including group size $g$, accumulation block size $B$, and tile sizes $N_{tile}$ and $K_{tile}$.
Specifically,}
\nickname can directly calculate the optimal tile size $K_{tile}$ and $N_{tile}$ based on hardware specifications, ensuring its out-of-box efficiency without tuning.

\paragraph{({\romannumeral 1}) Determine $N_{tile}$ by SIMD width.}
The gain of $N$ tiling comes from a reduced memory footprint while being able to utilize hardware parallelism (e.g., SIMD addition).
Therefore, it is optimal to minimize $N_{tile}$ while keeping it as multiples of the maximum SIMD width in INT16 (the accumulation precision).
For example, ARM NEON (128-bit)~\cite{ARM2025NEON} requires $8 \mid N_{tile}$, and AVX2 (256-bit)~\cite{Intel2022AVX2} requires $16 \mid N_{tile}$.
In our evaluation, $N_{tile}=32$ is empirically the most performant and universal configuration.

\paragraph{({\romannumeral 2}) Determine $K_{tile}$ by L1 cache size.}
The gain of $K$ tiling comes from reusing output registers when accumulating from different LUT channels (i.e., $K/g$) to the same output row (i.e., $M$), while not exceeding L1 cache size.
Therefore, it is optimal to maximize $K_{tile}$ below the cache-decided threshold, i.e., $3^gN_{tile}K_{tile}/g < \text{L1}$ for INT16 accumulation.
We use $K_{tile}=2g=10$ when $g=5$ and $K_{tile}=4g=16$ when $g=4$ during evaluation on most devices, except the AWS Graviton 3 server that performs best with $K_{tile}=g$.

\section{Evaluation}

We conduct a comprehensive evaluation across 3 real-world ternary LLMs and 5 devices, in both single and multi-thread.
The key takeaways are as follows:

\scalebox{0.8}{$\bullet$} \nickname outperforms all baselines (both LUT-based and MAD-based) on tested devices, with 1.5--5.0$\times$, 1.2--4.1$\times$, and 1.4--3.0$\times$ average speedup in the mpGeMM benchmark, end-to-end prefilling, and parallel decoding, respectively.

\scalebox{0.8}{$\bullet$} \nickname's \texttt{I1} (b1.60), which is the most compact ternary packing to date, outperforms all sub-2-bit baselines by up to $4.7\times$, $4.2\times$, and $3.2\times$, in the mpGeMM benchmark, end-to-end prefilling, and parallel decoding, respectively.

\scalebox{0.8}{$\bullet$} \nickname can seamlessly support mixed prefilling and decoding workloads like in continuous batching, achieving a 273.5 tokens/s throughput ($1.4\times$ of llama.cpp) on an affordable CPU server (\$0.50/h on AWS).

\scalebox{0.8}{$\bullet$} \revise{\nickname on CPU can match or surpass NPU inference performance, achieving 1.1$\times$ speedup over llama.cpp's Hexagon backend using only 2 CPU cores on Snapdragon 8 Elite.}

\scalebox{0.8}{$\bullet$} \revise{\nickname optimizes not just inference speed, but also energy efficiency, improving tokens/Joule by up to 2.1$\times$ and 1.8$\times$ over llama.cpp and T-MAC, respectively.}

\begin{table}[t]\small
  \caption{The support matrix of models, frameworks, and packings. llama.cpp falls back to 4-bit on HF BitNet 3B. LUT-based packings are underlined.}
  \setlength{\tabcolsep}{4pt} 
  \begin{tabular}{cccccc}
    \toprule
    \makecell{\textbf{Frame-}\\\textbf{works}} & \makecell{\textbf{Packing}\\\textbf{Methods}} & \textbf{BPWs} & \makecell{\textbf{HF BitNet}\\\textbf{3B~\cite{1bitllm2024bitnet}}} & \makecell{\textbf{Llama3}\\\textbf{8B~\cite{hf1bitllm2024llama3}}} & \makecell{\textbf{Falcon3}\\\textbf{1B~\cite{Falcon3}}} \\
    \midrule
    \multirow{2}{*}{Ours} & \underline{\textbf{\texttt{I2}}} & 2.00 & $\checkmark$ & $\checkmark$ & $\checkmark$ \\
    & \underline{\textbf{\texttt{I1}}} & 1.60 & $\checkmark$ & $\checkmark$ & $\checkmark$ \\
    \midrule
    T-MAC & \underline{\textbf{\texttt{INT\_N}}} & 2.00 & $\checkmark$ & $\checkmark$ & $\times$ \\
    \midrule
    \multirow{2}{*}{bitnet.cpp} & \texttt{I2\_S} & 2.00 & $\times$ & $\checkmark$ & $\checkmark$ \\
    & \underline{\textbf{\texttt{TL2}}} & 1.67 & $\checkmark$ & $\checkmark$ & $\times$ \\
    \midrule
    \multirow{2}{*}{llama.cpp} & \texttt{TQ2\_0} & 2.06 & \texttt{Q4\_0} & $\checkmark$ & $\checkmark$ \\
    & \texttt{TQ1\_0} & 1.69 & \texttt{Q4\_0} & $\checkmark$ & $\checkmark$ \\
    \bottomrule
  \end{tabular}
  \label{tab:model}
\end{table}

\begin{table*}[t]\small
  \caption{Specifications of tested devices. We only use performance cores for evaluation.}
  \label{tab:device}
  \begin{tabular}{lrrrr}
    \toprule
    \textbf{Device Name (Type)} & \textbf{Chip/Processor Name}  & \textbf{ISA \& SIMD}  & \textbf{Tested Cores}  & \textbf{L1D Cache per Core}\\
    \midrule
    Intel PC                    & Intel Core i7-13700k      & x86 AVX2 (256-bit)    & 4             & 48 KiB            \\ 
    Legion 5 Pro (Laptop)       & AMD Ryzen 7 5800H         & x86 AVX2 (256-bit)    & 4             & 32 KiB            \\ 
    Orange Pi 5 Plus (SBC)      & RK3588 (ARM Cortex-A76)   & ARM NEON (128-bit)    & 4             & 64 KiB            \\
    Xiaomi 15 (Smartphone)      & QCOM Snapdragon 8 Elite   & ARM NEON (128-bit)    & 2             & 96 KiB            \\
    AWS Graviton 3 (Server)     & ARM Neoverse-V1           & ARM SVE (256-bit)     & 8             & 64 KiB            \\
  \bottomrule
  \end{tabular}
\end{table*}

\begin{figure*}[t]
    \centering
    \newcommand{\subfigwidth}{\linewidth}
    \begin{subfigure}[b]{\subfigwidth}
        \centering
        \includegraphics[trim=0 10 0 5, clip, width=\linewidth]{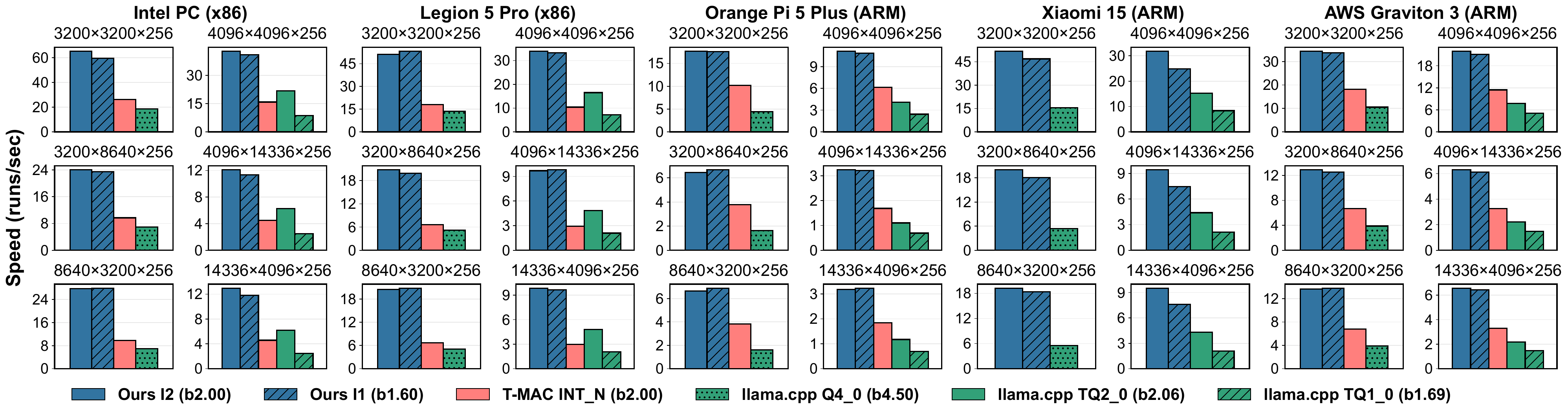}
        \caption{Single-thread mpGeMM}
    \end{subfigure}
    \hfill
    \begin{subfigure}[b]{\subfigwidth}
        \centering
        \includegraphics[trim=0 10 0 5, clip, width=\linewidth]{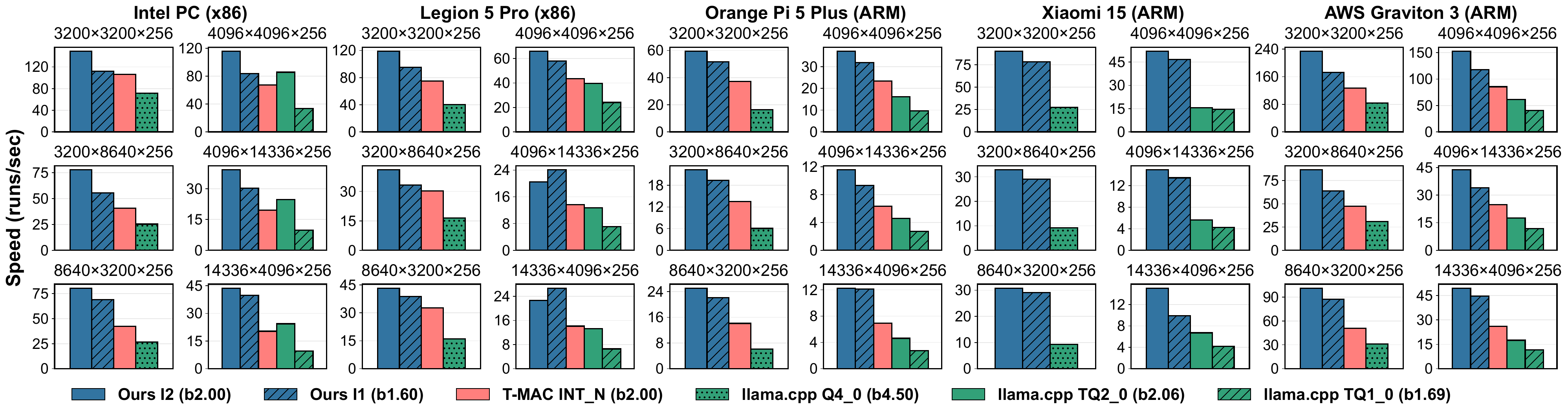}
        \caption{Multi-thread mpGeMM}
    \end{subfigure}
    \caption{mpGeMM kernel benchmark across devices and threads, using real-world LLMs' GeMM shapes.}
    \label{fig:eval-gemm}
\end{figure*}

\begin{figure*}[t]
    \centering
    \newcommand{\subfigwidth}{\linewidth}
    \begin{subfigure}[b]{\subfigwidth}
        \centering
        \includegraphics[trim=0 10 0 5, clip, width=\linewidth]{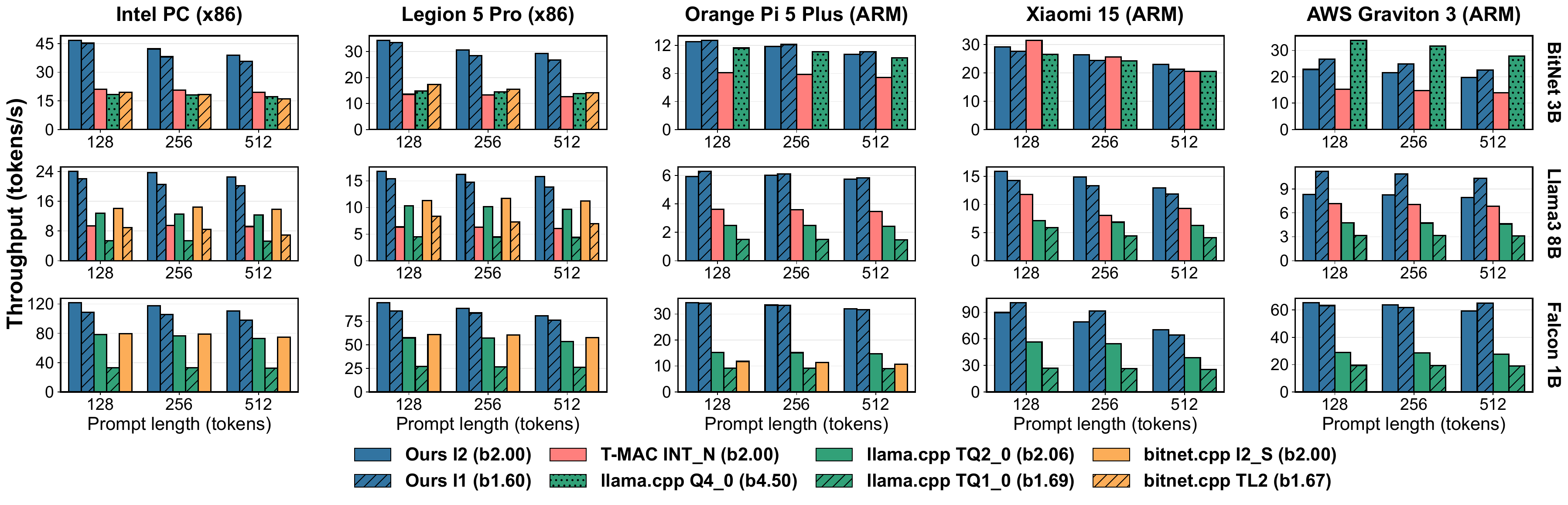}
        \caption{Single-thread prefilling}
    \end{subfigure}
    \hfill
    \begin{subfigure}[b]{\subfigwidth}
        \centering
        \includegraphics[trim=0 10 0 5, clip, width=\linewidth]{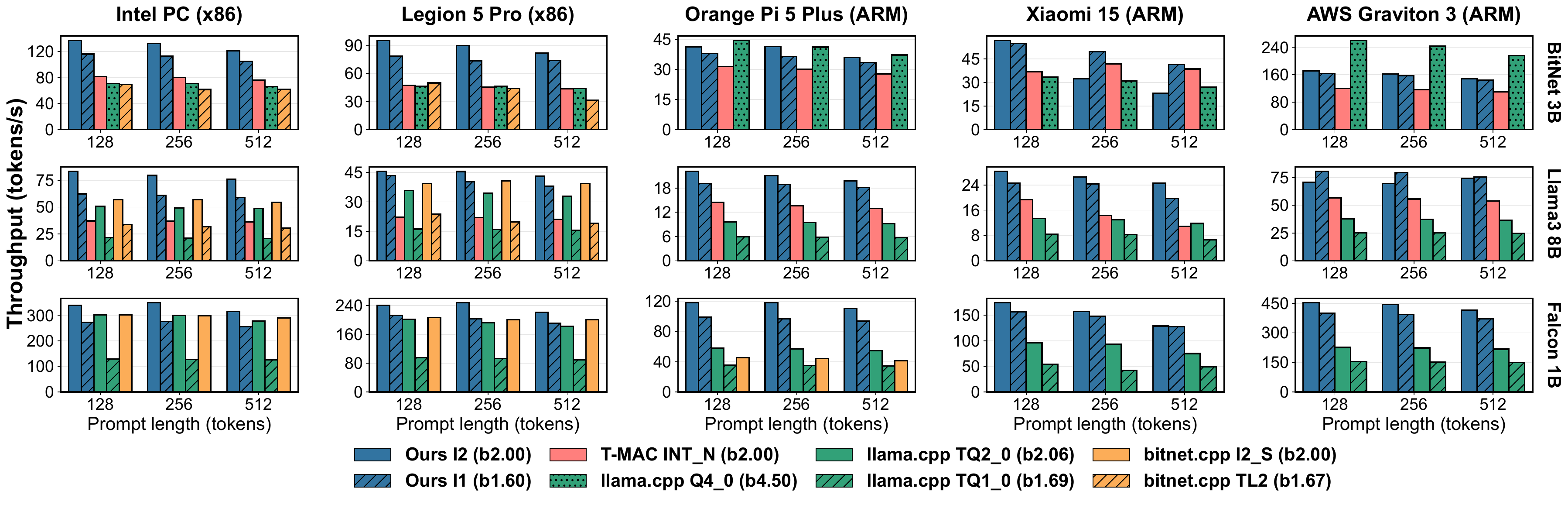}
        \caption{Multi-thread prefilling}
    \end{subfigure}
    \caption{End-to-end prefilling comparison across models, devices and threads. Sub-2-bit packings are hatched.}
    \label{fig:eval-prefill}
\end{figure*}

\subsection{Experimental Setup}

\paragraph{Devices}
We evaluate \nickname on 5 edge devices, as listed in Table~\ref{tab:device}.
We evaluate both single-thread and multi-thread inference, and only use performance cores since efficient cores would cause performance degradation for all frameworks in multi-thread inference.

\paragraph{Baseline frameworks and packings}
We compare \nickname with 3 baseline frameworks~\footnote{\revise{Unless otherwise noted, we use llama.cpp (\#833e2b7), T-MAC (\#8f29b96), and bitnet.cpp (\#8f75f99) in our evaluation; llama.cpp (\#5748be5), with Hexagon NPU support, is used specifically for a comparison with NPU (\S\ref{sec:eval-npu}).}} and 5 packings as listed in Table~\ref{tab:model}.
({\romannumeral 1}) llama.cpp~\cite{llamacpp} is the state-of-the-art MAD-based inference framework for ternary LLMs. It has two MAD-based packings: \texttt{TQ1\_0} (b1.69) and \texttt{TQ2\_0} (b2.06).
({\romannumeral 2}) T-MAC~\cite{wei2024tmac} is the state-of-the-art LUT-based mpGeMM kernel for ternary LLMs. It uses TVM~\cite{chen2018tvm} for kernel tuning, and is integrated with llama.cpp for end-to-end inference. It has one LUT-based packing: \texttt{INT\_N} (b2.00).
({\romannumeral 3}) bitnet.cpp~\cite{wang2025bitnet} is Microsoft's official inference framework for ternary LLMs, built atop llama.cpp. Its LUT-based kernel is based on T-MAC. It has 3 packings supporting different devices and models: MAD-based \texttt{I2\_S} (b2.00), and LUT-based \texttt{TL1} (b2.00) and \texttt{TL2} (b1.67).
We fail to run bitnet.cpp (\#8f75f99) on tested ARM devices, and only test it on x86 devices with supported packings.

\paragraph{Models and quantizations}
We use 3 real-world ternary LLMs to evaluate \nickname. The models and packings that support them are listed in Table~\ref{tab:model}.
We include both 2-bit and sub-2-bit packing for comparison as long as supported by models, frameworks, and devices.
One exception is that llama.cpp falls back to \texttt{Q4\_0} (b4.50) quantization for HF BitNet 3B, as it only supports weight shapes divisible by 256.
Although not our baseline for comparison due to the large gap in BPW ($>2\times$), we still show its evaluation results.

\revise{\paragraph{Tasks and metrics}
The evaluation is focused on system performance, rather than model accuracy, since our method is lossless for ternary weights.
The main evaluation includes a kernel benchmark and 3 end-to-end comparisons.
In \S\ref{sec:eval-gemm}, we compare operator throughput (runs/s, higher is better) of the mpGeMM kernel.
In \S\ref{sec:eval-e2e}, we compare token throughput (tokens/s, higher is better) on 3 real-world workloads, including prefilling (\S\ref{sec:eval-prefill}), parallel decoding (\S\ref{sec:eval-batch}), and continuous batching (\S\ref{sec:eval-batch}) that dynamically mixes prefilling and decoding.
Additionally, we compare energy efficiency (tokens/Joule, higher is better) in \S\ref{sec:eval-energy}.}

\subsection{mpGeMM Kernel Benchmark}

\label{sec:eval-gemm}

We evaluate the mpGeMM kernel performance on different devices in different threads, using real-model GeMM shapes (results in Fig.~\ref{fig:eval-gemm}).
T-MAC is tuned for tested GeMM shapes.
bitnet.cpp is excluded as it doesn't provide a kernel-level benchmark tool, and is similar to T-MAC's design.

\paragraph{Comparison with MAD-based mpGeMM}
Both \texttt{I1} and \texttt{I2} packing of \nickname outperform MAD-based llama.cpp in all test cases.
When comparing in similar BPWs (i.e., \texttt{I1} (b1.60) vs. \texttt{TQ1\_0} (b1.69) and \texttt{I2} (b2.00) vs. \texttt{TQ2\_0} (b2.06)), \nickname respectively achieve 3.4--4.7$\times$ and 1.7--2.9$\times$ speedup in single-thread mpGeMM, and 2.9--3.7$\times$ and 1.6--2.8$\times$ speedup in multi-thread mpGeMM, averaged among GeMM shapes.
Notably, \nickname's \texttt{I1} achieves comparable or slightly higher performance than \texttt{I2}, while llama.cpp's \texttt{TQ1\_0} is significantly slower than \texttt{TQ2\_0}, which demonstrates the bit-scaling efficiency of \nickname.

\paragraph{Comparison with scalar LUT-based mpGeMM}
\nickname also outperforms scalar LUT-based T-MAC in all test cases.
Specifically, \nickname's \texttt{I2} (b2.00) achieves 1.8--3.2$\times$ and 1.5--1.9$\times$ average speedup over T-MAC's \texttt{INT\_N} (b2.00) in single-thread and multi-thread mpGeMM, respectively.
In some cases (e.g., Intel PC and AMD laptop), T-MAC cannot outperform llama.cpp's MAD-based \texttt{TQ2\_0}, which proves the necessity of our vector LUT design for parallel inference.

\paragraph{Speedup with different threads}
Although \nickname outperforms baselines in different threads, the improvement in multi-thread is less significant than in single-thread.
For example, the speedup of \nickname's \texttt{I1} over llama.cpp's \texttt{TQ1\_0} drops from 3.4--4.7$\times$ in single-thread to 2.9--3.7$\times$ in multi-thread.
The main reason is that during LUT precomputing, different threads access the activation concurrently, and compete for the main memory bandwidth.

\subsection{End-to-End LLM Inference}

\label{sec:eval-e2e}
We evaluate the end-to-end inference throughput on different devices, threads, and workloads (prefilling, parallel decoding, and continuous batching).
We follow llama.cpp's practice to quantize output and token embedding weights to \texttt{Q6\_K} and \texttt{Q4\_K}, respectively~\cite{compilade2024tq}.
We denote llama.cpp's \texttt{TQ2\_0} and \texttt{TQ1\_0} as \texttt{Q4\_0} explicitly when they fall back on unsupported GeMM shapes.
bitnet.cpp is only evaluated on x86 devices since it fails to run on the tested ARM devices.

\subsubsection{Prefilling}

\label{sec:eval-prefill}

The prefilling evaluation covers all devices, frameworks, and models, with different prompt lengths and thread configurations (results in Fig.~\ref{fig:eval-prefill}).

\paragraph{Comparison with MAD-based inference}
There are 3 MAD-based baselines: llama.cpp's \texttt{TQ1\_0} (b1.69) and \texttt{TQ2\_0} (b2.06), and bitnet.cpp's \texttt{I2\_S} (2.00).
\nickname's \texttt{I2} (b2.00) outperforms them consistently on different devices, by 3.0--4.1$\times$, 1.5--2.9$\times$, and 1.6--2.3$\times$ in single-thread, and 2.7--3.5$\times$, 1.2--2.6$\times$, and 1.2--2.2$\times$ in multi-thread, averaged among different models and prompt lengths.
For sub-2-bit packing specifically, \nickname's \texttt{I1} (b1.60) outperforms llama.cpp's \texttt{TQ1\_0} by 3.1--4.0$\times$ in single-thread and 2.5--3.0$\times$ in multi-thread.
Despite the overall speedup (consistent with \S\ref{sec:eval-gemm}), llama.cpp's \texttt{Q4\_0} on AWS Graviton 3 is an exception.
The main reason is that llama.cpp's 4-bit kernel is highly optimized on ARM devices with hardware intrinsics, outperforming both its 2-bit kernel and T-MAC.
Considering the large gap in memory requirements (by $>2\times$), \nickname is still preferable in memory-constrained scenarios.

\paragraph{Comparison with scalar LUT-based inference}
There are 2 LUT-based baselines: T-MAC's \texttt{INT\_N} (b2.00) and bitnet.cpp's \texttt{TL2} (b1.67).
\nickname outperforms both of them, except HF BitNet 3B prefilling on the smartphone in single-thread.
When comparing in similar BPWs (i.e., \nickname's \texttt{I2} (b2.00) vs. T-MAC, and \texttt{I1} (b1.67) vs. bitnet.cpp's \texttt{TL2}), \nickname respectively achieves 1.1--2.6$\times$ and 1.9--2.4$\times$ single-thread speedup, and 1.3--2.0$\times$ and 1.7--1.8$\times$ multi-thread speedup.

\subsubsection{Parallel Decoding and Continuous Batching}

\label{sec:eval-batch}

\begin{figure}[t]
    \centering
    \includegraphics[trim=0 10 0 5, clip, width=\linewidth]{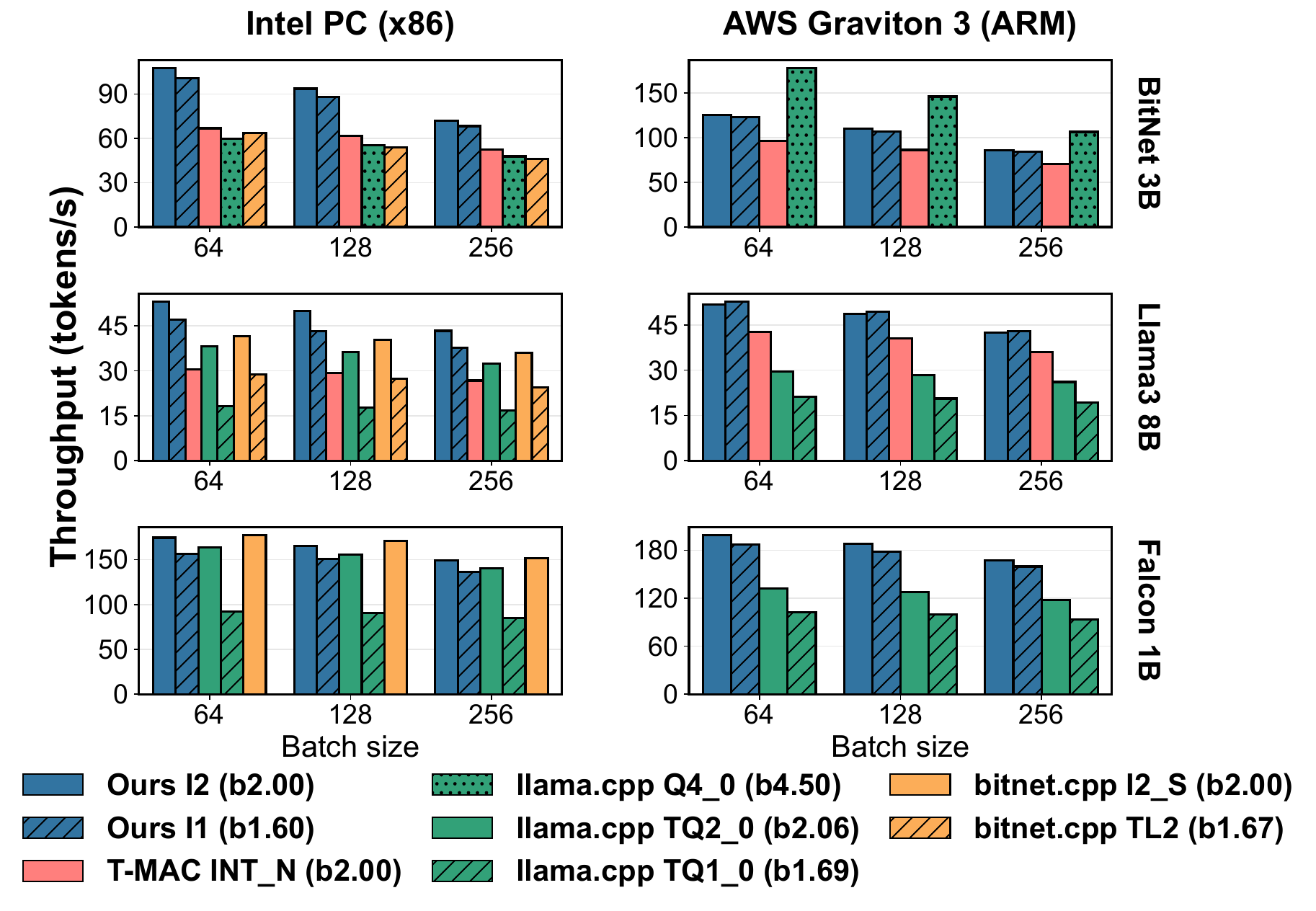}
    \caption{Multi-thread parallel decoding comparison across models and devices. Sub-2-bit packings hatched.}
    \label{fig:eval-batch}
\end{figure}

The parallel decoding evaluation is conducted on the PC and server in multi-thread, to reflect real-world deployment configurations.
We measure the decoding throughput with a fixed prompt length and token generation length (16), and varying batch sizes (results in Fig.~\ref{fig:eval-batch}).
We also evaluate continuous batching, to demonstrate \nickname's capability to support more complex parallel workloads with mixed prefilling and decoding.

\paragraph{Parallel decoding}
\nickname achieves similar speedup in parallel decoding as in prefilling, with up to $2.1\times$, $2.1\times$, and $3.0\times$ improvements over T-MAC, bitnet.cpp, and llama.cpp, respectively.
Despite the overall improvements, we notice 2 exceptions.
(\romannumeral 1) llama.cpp's \texttt{Q4\_0} still outperforms \nickname for HF BitNet 3B on AWS Graviton 3, due to the same reason explained in \S\ref{sec:eval-prefill}.
(\romannumeral 2) bitnet.cpp's MAD-based \texttt{I2\_S} (b2.00) slightly outperforms \nickname on Intel PC and AMD laptop with Falcon 1B. The main reason is that Falcon 1B has much smaller $M$ in its GeMM shapes (e.g., $2048-6144$ in Falcon 1B vs. $4096-14436$ in Llama3 8B), which limits the benefit from table lookup\revise{, for both vector and scalar LUT}.
\revise{In practice, the switching point can be determined by profiling.}
Besides, the concurrent memory access, as discussed in \S\ref{sec:eval-gemm}, also makes \nickname's multi-thread acceleration less significant.

\paragraph{Continuous Batching}
We conduct an end-to-end serving test with continuous batching to showcase \nickname's capability to support complex parallel workloads\revise{ with mixed prefilling and decoding in dynamic batches}.
On an 8-core AWS Graviton 3 server that costs only \$0.50/h, \nickname's \texttt{I2} can serve Falcon 1B for a 273.5 tokens/s average throughput over 32 parallel requests, including 152.6 tokens/s prefilling and 120.9 tokens/s decoding, outperforming llama.cpp's \texttt{TQ2\_0} by $1.4\times$.
This also demonstrates the viability of deploying ternary LLMs on affordable CPU servers, making advanced language models more accessible without requiring specialized hardware or expensive computational resources.

\begin{figure}[t]
    \centering
    \includegraphics[trim=0 10 0 5, clip, width=\linewidth]{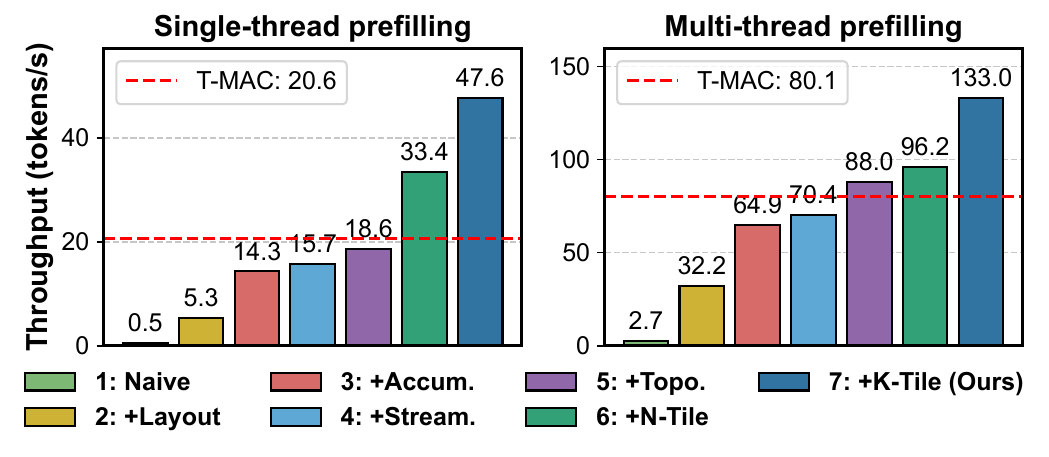}
    \caption{Prefilling throughput of HF BitNet 3B on Intel PC, with \nickname's techniques applied step by step. Accum: hierarchical accumulation; Stream: streamed lookup; Topo: topological precomputing.}
    \label{fig:eval-ablation}
\end{figure}

\begin{figure}[t]
    \centering
    \includegraphics[trim=0 0 0 0, clip, width=0.8\linewidth]{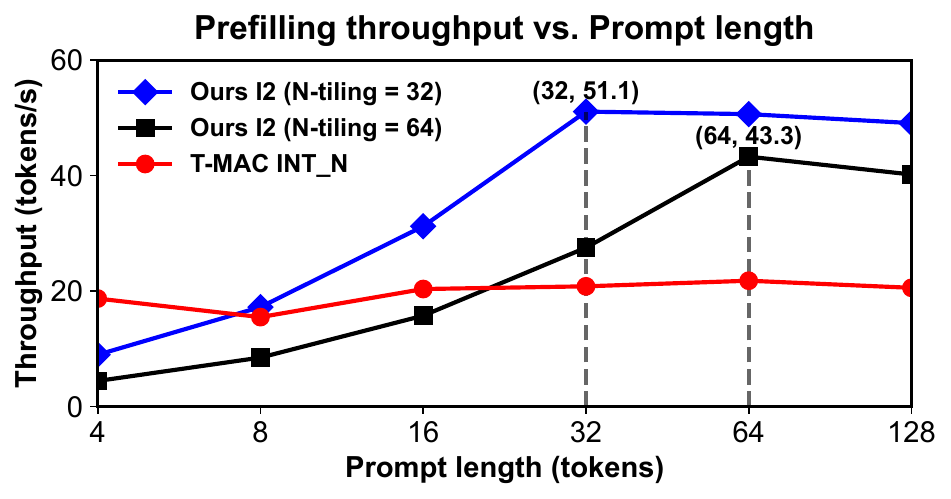}
    \caption{Prefilling throughput of HF BitNet 3B on Intel PC, with different prompt lengths. As the prompt length increases, \nickname's throughput improves, but T-MAC's throughput remains nearly constant.}
    \label{fig:eval-prefill-length}
\end{figure}

\begin{table}[t]\small
  \caption{\nickname's latency breakdown in HF BitNet 3B prefilling. Measured with VTune~\cite{Intel2025VTune}.}
  \setlength{\tabcolsep}{4pt}
  \begin{tabular}{lrrrr}
    \toprule
    Total & Precomp. (\%) & Accum. (\%) & \textbf{Lookup (\%)} & Others (\%)\\
    \midrule
    15.05s & 1.85 & 65.24 & \underline{\textbf{0.59}} & 32.32 \\
  \bottomrule
  \end{tabular}
  \label{tab:breakdown}
\end{table}


\begin{table}[t]\small
\centering
\caption{\revise{Energy efficiency (tokens/Joule) on AMD Ryzen 7 5800H with Llama3 8B (prefilling 256 tokens): \nickname (\texttt{I2}) vs.\ llama.cpp (\texttt{TQ2\_0}) and T-MAC (\texttt{INT\_N}).}}
\label{tab:energy}
\setlength{\tabcolsep}{4pt}
\begin{tabular}{cccc}
\toprule
\textbf{\#Threads} & \textbf{llama.cpp} & \textbf{T-MAC} & \textbf{\nickname} \\
\midrule
1 & 0.80 & 1.13 & \underline{\textbf{1.70}} \\
4 & 1.32 & 1.40 & \underline{\textbf{2.54}} \\
\bottomrule
\end{tabular}
\end{table}

\begin{table}[t]\small
\centering
\caption{\revise{Throughput (tokens/s) on Snapdragon 8 Elite with Llama3 8B
(prefilling 32/128 tokens): \nickname (\texttt{I2}, CPU) vs.\ llama.cpp's
Hexagon backend (\texttt{Q4\_0}, NPU).}}
\label{tab:npu}
\setlength{\tabcolsep}{4pt}
\begin{tabular}{cccc}
\toprule
\textbf{\#Tokens} & \textbf{llama.cpp} & \textbf{\nickname} & \textbf{\nickname} \\
 & \textbf{(NPU)} & \textbf{(single-thread)} & \textbf{(dual-thread)} \\
\midrule
32 & 26.79 & 19.38 & \underline{\textbf{30.11}} \\
128 & 24.89 & 18.46 & \underline{\textbf{26.05}} \\
\bottomrule
\end{tabular}
\end{table}

\subsection{Breakdown Analysis}

\label{sec:eval-breakdown}

Table~\ref{tab:breakdown} shows the end-to-end prefilling latency breakdown of \nickname.
Non-mpGeMM operators (e.g., floating-point GeMM in Attention) take up 32\% of the total latency,
results accumulation (essentially vector addition) takes up 65\% of the total latency, and precomputing and lookup consume little execution time.
Notably, comparing to T-MAC (results in Table~\ref{tab:tmac-breakdown}), the ratio of lookup cost drops from 47\% to below 1\%, and the ratio of accumulation increases from 25\% to 64\%.
This validates our vector LUT paradigm that turns random and non-contiguous table lookup into contiguous vector addition, and could inspire future works on vector LUT-centric accelerator design.

\subsection{Ablation Study}

\label{sec:eval-ablation}

\paragraph{Different techniques}
Fig.~\ref{fig:eval-ablation} \revise{isolates the benefits of} all main techniques of \nickname, including vector LUT-centric tensor layout, hierarchical accumulation, streamed precomputing-lookup, topological precomputing, and locality-aware tiling.
Specifically, the layout optimizations provide the initial 10.6--11.9$\times$ speedup, and the streamed LUT optimizations provide another 4.1--9.0$\times$ speedup.

\paragraph{Different input lengths and tile sizes}
Fig.~\ref{fig:eval-prefill-length} shows \nickname's performance under different input lengths and $N$-tiling sizes.
Unlike T-MAC's constant throughput despite increased parallel tokens,
\nickname's throughput increases linearly with the prompt length, \revise{surpassing T-MAC at $N=8$ ($N_{tile}=32$),}and reaches its peak when the prompt length equals $N$-tiling size.
Also, the tile size of 32 outperforms 64, consistent with the analysis in \S\ref{sec:impl}.

\revise{

\subsection{Energy Efficiency}

\label{sec:eval-energy}

Table~\ref{tab:energy} compares the energy efficiency of \nickname and baselines in both single-thread and multi-thread prefilling.
\nickname improves tokens/Joule by 1.9--2.1$\times$ over llama.cpp and by 1.5--1.8$\times$ over T-MAC, demonstrating substantial energy savings.
These improvements stem from ({\romannumeral 1}) LUT-based mpGeMM that eliminates costly dequantization and multiplication (vs. MAD-based llama.cpp), ({\romannumeral 2}) an efficient memory access pattern that reduces DRAM energy cost (vs. scalar LUT-based T-MAC), and ({\romannumeral 3}) higher throughput that amortizes fixed power overhead across more tokens per second.
}

\revise{

\subsection{\revise{Comparison with NPU}}

\label{sec:eval-npu}

Table~\ref{tab:npu} compares the throughput of \nickname (\texttt{I2}, CPU) with llama.cpp's Hexagon backend (\texttt{Q4\_0}, NPU) on Snapdragon 8 Elite~\footnote{The Hexagon backend of llama.cpp (\#833e2b7) doesn't support fewer bits than 4.}.
With a single CPU core, \nickname achieves 72\%--74\% of the NPU throughput; with two CPU cores, it reaches 1.05--1.12$\times$, despite Snapdragon 8 Elite having a powerful NPU.
By leveraging efficient LUT memory access, \nickname enables CPU-based inference to match or surpass NPU-based inference on the same device, providing deployment flexibility without requiring NPU support.
}


\section{Discussion}

\subsection{Different Hardware Platforms}
Following existing LUT-based works~\cite{wei2024tmac,wang2025bitnet}, we demonstrate \nickname's effectiveness on the ubiquitous and LUT-friendly CPU. It is particularly important for edge and IoT devices. 
Specifically, CPU has a good support for table lookup and vector addition (\nickname's main operations).
Other hardware (e.g., GPU and NPU), although becoming prevalent on new-generation devices, are multiplication-centric, and don't suit ultra-low-bit inference.
Besides, it is often impractical to use them (e.g., GPU might be reserved for rendering only), or difficult to program them (e.g., the NPU ecosystem is quite closed and tends to support fixed quantization formats).
Moreover, we hope \nickname will inspire future research on vector LUT-centric hardware.

\subsection{Different Bits of Quantization}
\label{sec:discuss-bits}

Current \nickname implementation particularly targets the Pareto-optimal ternary LLMs. The idea of vector LUT could be extended to different bits (e.g., 1, 2, 3, 4-bit) by adapting the weight packing and indexing methods (e.g., in Fig.~\ref{fig:bit-map}).
Besides, \nickname's design is orthogonal to T-MAC's bit decomposition technique, which decomposes n-bit weights to n of 1-bit weights for unified handling of different quantization bits.
The main trade-off is that higher bits will increase the LUT size and lookup complexity.

\subsection{Combination with Existing Methods}
\revise{
\nickname is primarily optimized for tasks with long inputs and short outputs (e.g., summarization) or parallel generation.
For more diverse workloads, \nickname can be combined with scalar LUT-based methods, switching between them based on the input token length to achieve end-to-end speedup (trade-off shown in Fig.~\ref{fig:eval-prefill-length}).
}
For example, when processing a single user request, it is optimal to use \nickname during prefilling (parallel), and switch to scalar LUT-based kernels during decoding (usually not parallel), to minimize the end-to-end inference latency.
\revise{
Switching between vector and scalar LUT requires weight re-layout due to their different tiling strategies, incurring extra latency if converted at runtime or extra memory if pre-converted.
}
\section{Related Works}

\subsection{Ultra-Low-Bit LLM Quantization}
\label{sec:related-quant}

Large language models (LLMs) have demonstrated remarkable capabilities but require substantial resources for deployment.
To mitigate this, low-bit quantization has emerged as a crucial technique to reduce model size while maintaining model accuracy.
For example, LLM.int8()~\cite{dettmers2022gpt3} and SmoothQuant~\cite{xiao2023smoothquant} adopt 8-bit quantization for both weights and activations to save memory and computation.
GPTQ~\cite{frantar2022gptq} and AWQ~\cite{lin2023awq} further compress the LLM weights to 4 and 3-bit with comparable accuracy to floating point baselines, while keeping activations at FP16 due to random outliers.
Besides the post-training quantization (PTQ) methods above,
quantization-aware training (QAT) methods (e.g., BitDistiller~\cite{du2024bitdistiller}, EfficientQAT~\cite{chen2025efficientqat}, and ParetoQ~\cite{liu2025paretoq}) further push the bit width boundary to 2-bit and below.
Moreover, Microsoft trains the native ternary (1.58-bit) LLM BitNet from scratch~\cite{wang2023bitnet,ma2024158bitllm,ma2025bitnet}, which is followed by the community with many emerging ternary LLMs~\cite{1bitllm2024bitnet,hf1bitllm2024llama3,Falcon3,kaushal2024spectra}.
However, the mixed-precision challenge of low-bit, especially ternary LLMs, remains unsolved and significantly hinders their practical deployment on current hardware.

\subsection{LUT-based mpGeMM Optimization}
To efficiently support low-bit LLM inference without native hardware support of mpGeMM, LUT-based methods are proposed in replacement of conventional MAD-based methods (more in \S\ref{sec:bg-mpgemm}).
For example, LUT-GEMM~\cite{park2022lut} first proposes LUT-based mpGeMM for low-bit LLMs on GPU.
T-MAC~\cite{wei2024tmac} utilizes the table lookup instructions on CPU to accelerate LUT-based mpGeMM, and even outperforms GPU/NPU-based solutions in single-batch decoding.
bitnet.cpp~\cite{wang2025bitnet}, Microsoft's official inference framework for ternary LLMs, extends T-MAC's LUT-based kernel design with ternary-specific optimizations and achieves a remarkable inference speed on CPU.
Architectural works (e.g., LUT Tensor Core~\cite{mo2024lut}, TENET~\cite{huang2025tenet}, and T-SAR~\cite{oh2025tsar}) further accelerate LUT-based mpGeMM by co-designing the LUT algorithm and hardware architecture, demonstrating the superior efficiency of LUT-based solutions.
However, all of the existing methods adopt the scalar LUT paradigm, which is inefficient for parallel inference, as discussed in \S\ref{sec:bg-lut-limit}.
\section{Conclusion}
We present \nickname, a LUT-based mpGeMM kernel for parallel inference of ultra-low-bit LLMs on edge devices.
Through the novel vector LUT paradigm that performs $1\rightarrow N$ lookup, instead of $1\rightarrow 1$ lookup in existing scalar LUT-based kernels,
\nickname avoids repetitive table loading and lookup, and accelerates mpGeMM via fully utilizing the memory bandwidth. 
On the ubiquitous CPU backend with llama.cpp integration, \nickname's \texttt{I1} (b1.60) and \texttt{I2} (b2.00) achieve up to $4.2\times$ and $2.6\times$ prefilling speedup respectively, compared to baselines of similar BPWs.
Moreover, \nickname's \texttt{I1} is the most compact and versatile ternary packing for now.

\begin{acks}
  This work is supported by NSFC (No.62402280), CPSF (No.2024M761683), Tsinghua University (AIR)-AsiaInfo Technologies (China), Inc. Joint Research Center for 6G Network and Intelligent Computing, and Xiongan AI Institute.
\end{acks}

\bibliographystyle{ACM-Reference-Format}
\bibliography{ref}









\appendix
\section{Artifact Appendix}

\subsection{Abstract}

The source code for this work is \reponame, available at \codelink.
Pre-converted GGUF models are hosted on HuggingFace at \hflink.
Detailed evaluation instructions (\texttt{Evaluation.md}) and all scripts are in the \texttt{evaluation/} directory of the repository.

The instructions below focus on a minimal reproduction of \reponame on a single device with one pre-converted model ($<$1\,hour).
Extending to other models and devices is straightforward (same scripts, different \texttt{MODEL\_DIR} and \texttt{DEVICE\_NAME}).
A full comparison against all baselines requires additional setup described in \texttt{Evaluation.md} ($>$1\,day per device).

\subsection{Artifact check-list (meta-information)}

{\small
  \begin{itemize}
    \item {\bf Algorithm:} vector LUT-based mpGeMM
    \item {\bf Compilation:} CMake $\geq$3.14, C++17 (GCC $\geq$9 / Clang $\geq$10)
    \item {\bf Model:} pre-converted GGUF models, available at \hflink
    \item {\bf Run-time environment:} Ubuntu 20.04+ / WSL2 / Android
    \item {\bf Hardware:} x86\_64 AVX2 or ARMv8 NEON/SVE CPU; $\geq$4\,GB RAM (inference only) or $\geq$24\,GB RAM (manual model conversion)
    \item {\bf Execution:} exclusive access to performance cores
    \item {\bf Metrics:} kernel throughput (runs/s), token throughput (tokens/s)
    \item {\bf Output:} CSV summaries, PDF figures via Python plotting
    \item {\bf Experiments:} GeMM benchmark, end-to-end prefilling, end-to-end batched decoding
    \item {\bf How much disk space required (approximately)?:} $\geq$16\,GB for \reponame models; $\geq$64\,GB for full comparison with baselines
    \item {\bf How much time is needed to prepare workflow (approximately)?:} $\sim$30\,min (build + model download)
    \item {\bf How much time is needed to complete experiments (approximately)?:} $<$1\,hour for minimal reproduction; $>$1\,day per device for full comparison
    \item {\bf Publicly available?:} Yes
    \item {\bf Code licenses (if publicly available)?:} MIT
  \end{itemize}
}

\subsection{Description}

\subsubsection{How to access}
\begin{itemize}
  \item Source code: \codelink
  \item Pre-converted models: \hflink
\end{itemize}

\subsubsection{Hardware dependencies}

\begin{itemize}
  \item Minimum: x86\_64 CPU with AVX2, or ARMv8 CPU with NEON/SVE, $\geq$4\,GB RAM
  \item Paper devices: Intel Core i7-13700K (PC), AMD Ryzen 7 5800H (laptop), RK3588/Cortex-A76 (Orange Pi 5 Plus SBC), Snapdragon 8 Elite (Xiaomi 15), ARM Neoverse-V1 (AWS Graviton 3)
\end{itemize}

\subsubsection{Software dependencies}

\begin{itemize}
  \item OS: Linux (Ubuntu 20.04+), Android (Termux), or WSL2
  \item Build: CMake $\geq$3.14, C++17 (GCC $\geq$9 or Clang $\geq$10)
  \item Python $\geq$3.10 with \texttt{huggingface\_hub}, \texttt{pandas}, \texttt{matplotlib}
\end{itemize}

\subsubsection{Models}

\label{sec:desc-model}

Pre-converted GGUF models for \reponame are hosted at \hflink:

\begin{itemize}
  \item HF BitNet 3B: \href{https://huggingface.co/XXXXyu/bitnet\_b1\_58-3B-vlut-gguf}{\texttt{XXXXyu/bitnet\_b1\_58-3B-vlut-gguf}}
  \item Llama3 8B: \href{https://huggingface.co/XXXXyu/Llama3-8B-1.58-100B-tokens-vlut-gguf}{\texttt{XXXXyu/Llama3-8B-1.58-100B-tokens-vlut-gguf}}
  \item Falcon3 1B: \href{https://huggingface.co/XXXXyu/Falcon3-1B-Instruct-1.58bit-vlut-gguf}{\texttt{XXXXyu/Falcon3-1B-Instruct-1.58bit-vlut-gguf}}
\end{itemize}

For manual model conversion or baseline evaluation, use the original HuggingFace weights (see \texttt{Evaluation.md}).

\subsection{Installation}

\subsubsection{Build from source.}
Clone the repository and build with CMake:

\begin{lstlisting}
git clone https://github.com/OpenBitSys/vlut.cpp.git
cd vlut.cpp
cmake -B build
cmake --build build --config Release -j4
\end{lstlisting}

Optional CMake flags: \texttt{-DVLUT\_SVE=ON} for ARM SVE (e.g., AWS Graviton 3); \texttt{-DTABLE\_ENTRY\_SIZE=<N>} to change the $N$-tile size (default: 32).

\subsubsection{Set up Python and download models.}
Install Python dependencies and download pre-converted GGUF models. Store each model in a directory named after the model's short name:

\begin{lstlisting}
pip install huggingface_hub pandas matplotlib
# download pre-converted models (example: Llama3 8B)
hf download XXXXyu/Llama3-8B-1.58-100B-tokens-vlut-gguf \
  ggml-model-I1_V_2.gguf ggml-model-I2_V_4.gguf \
  --local-dir ~/models/Llama3-8B-1.58-100B-tokens
\end{lstlisting}

Repeat for other models, replacing the repository ID and \texttt{-\-local-dir} accordingly.

\subsection{Experiment Workflow}

\subsubsection{Quick functional check.}
Verify the build by running a single batched-decoding pass:

\begin{lstlisting}
./build/bin/llama-batched \
  -m ~/models/Llama3-8B-1.58-100B-tokens/ggml-model-I1_V_2.gguf \
  -p "I believe" -np 32 -n 16 -t 4 -ngl 0 \
  --temp 0.5 --repeat-penalty 1.5
\end{lstlisting}

The model should load without errors and print generated text from 32 parallel sequences.

\subsubsection{Minimal reproduction (one device, one model).}
\label{sec:minimal}

Set environment variables and run the three benchmarks:

\begin{lstlisting}
export DEVICE_NAME=mydevice
export MODEL_DIR=~/models/Llama3-8B-1.58-100B-tokens

# 1. GeMM kernel benchmark (single- and multi-threaded)
./evaluation/scripts/bench-gemm.sh "$DEVICE_NAME" 1 256
./evaluation/scripts/bench-gemm.sh "$DEVICE_NAME" 4 256
# 2. End-to-end prefilling
DEVICE_NAME="$DEVICE_NAME" MODEL_DIR="$MODEL_DIR" \
  ./evaluation/scripts/bench-e2e-prefill.sh
# 3. End-to-end batched decoding
DEVICE_NAME="$DEVICE_NAME" MODEL_DIR="$MODEL_DIR" \
  ./evaluation/scripts/bench-e2e-batch.sh
\end{lstlisting}

Raw results (CSV/log) are written to directories under \texttt{evaluation/}, named \texttt{results\_gemm\_\allowbreak\$\{DEVICE\_NAME\}}, \texttt{results\_e2e\_\allowbreak prefill\_\allowbreak\$\{DEVICE\_NAME\}}, and \texttt{results\_e2e\_\allowbreak batch\_\allowbreak\$\{DEVICE\_NAME\}}.
The scripts benchmark all \reponame quantization variants found in \texttt{MODEL\_DIR} and skip missing baselines with warnings.

\subsubsection{Other models and devices.}
To evaluate additional models, change \texttt{MODEL\_DIR} and re-run Steps 2--3 (the GeMM benchmark uses built-in shapes and only needs to run once per device).
To evaluate additional devices, run on different hardware with a new \texttt{DEVICE\_NAME} (results saved with different folder names).

\subsubsection{Full paper comparison.}
The same scripts automatically benchmark baselines (\texttt{llama.cpp}, \texttt{T-MAC}, \texttt{bitnet.cpp}) when their repositories and model files are present.
Place all repositories as siblings in the same workspace and prepare baseline model files as described in \texttt{Evaluation.md}.
Note that \texttt{T-MAC} and \texttt{bitnet.cpp} may require re-compilation per model or quantization variant.

\subsection{Evaluation and Expected Results}

Generate figures and summary reports from raw results.
Each plotting script accepts \texttt{-{}-single-thread}, \texttt{-{}-multi-thread}, or \texttt{-{}-both} (default):

\begin{lstlisting}
python evaluation/scripts/plot/plot_gemm_combined.py
python evaluation/scripts/plot/plot_e2e_prefill_combined.py
python evaluation/scripts/plot/plot_e2e_batch_combined.py
\end{lstlisting}

Output PDF figures are saved to \texttt{evaluation/figures/} and CSV reports to \texttt{evaluation/reports\_*/}.
The device identifier \texttt{mydevice} is recognized by the plotting scripts as a built-in dummy identifier.
The paper's canonical identifiers are: \texttt{pc\_intel}, \texttt{laptop\_amd}, \texttt{orangepi}, \texttt{smartphone}, \texttt{aws\_arm}.

Absolute throughput varies by hardware, but relative speedup trends should match the paper:
\reponame generally outperforms baselines in GeMM throughput, prefilling speed, and parallel decoding throughput; the \texttt{I1} quantization remains competitive while using fewer bits.
On AWS Graviton 3, \texttt{llama.cpp}'s \texttt{Q4\_0} for HF BitNet 3B can be relatively strong due to ARM-specific 4-bit kernel optimizations.
Without baselines, CSV speedup reports will be incomplete---verify raw results and figures instead.

\subsection{Experiment Customization}

\subsubsection{Build-time options.}
\reponame exposes two CMake flags that affect performance:

\begin{itemize}
  \item \texttt{-DTABLE\_ENTRY\_SIZE=<N>}: $N$-tile size (default: 32). Larger values (e.g., 64) may improve throughput on some CPUs.
  \item \texttt{-DVLUT\_SVE=ON}: enable ARM SVE intrinsics (e.g., AWS Graviton 3).
\end{itemize}

\subsubsection{Quantization types.}
Five Vec-LUT packings are available (\texttt{I1\_V}, \texttt{I1\_V\_2}, \texttt{I2\_V}, \texttt{I2\_V\_4}, \texttt{I2\_V\_8}), controlling the $K$-tiling and packed weight layout. All five are provided as pre-converted models on HuggingFace. The examples above use \texttt{I1\_V\_2} and \texttt{I2\_V\_4}, which are performant on most devices. To convert models with a specific packing manually, see \texttt{Evaluation.md}.

\subsubsection{Benchmark script arguments.}
The GeMM script takes positional arguments: \texttt{bench-gemm.sh <device> <threads> <seq\_len> [entry\_size]}.
The end-to-end scripts accept environment variables to override defaults:

\begin{itemize}
  \item Prefilling (\texttt{bench-e2e-prefill.sh}): \texttt{PROMPT\_LENGTH} (default \texttt{128,256,512}), \texttt{THREAD\_COUNT} (default \texttt{1,4}), \texttt{REPEAT\_COUNT} (default \texttt{3}).
  \item Batched decoding (\texttt{bench-e2e-batch.sh}): \texttt{PARALLEL\_SEQS} (default \texttt{64,128,256}), \texttt{THREAD\_COUNT} (default \texttt{4}), \texttt{PREFILL\_LEN} (default \texttt{16}), \texttt{TOKEN\_GEN\_LENS} (default \texttt{16}).
\end{itemize}

\subsubsection{Automated configuration search.}
\texttt{search-config.sh} sweeps over $N$-tile sizes (16, 32, 64) and thread counts (1, 4, 8) to find the best configuration for a given device. Results are written to \texttt{evaluation/results\_search/}.

\subsubsection{Other models.}
Beyond the three models evaluated in the paper, \reponame also supports more ternary models. Convert and evaluate them following the instructions in \texttt{Evaluation.md}.

\subsection{Notes}

\begin{itemize}
  \item Set the CPU governor to \texttt{performance} mode and minimize background processes for stable results.
  \item On heterogeneous CPUs (e.g., ARM big.LITTLE), use \texttt{-{}-cpu-mask} and \texttt{-{}-cpu-strict} to pin inference to performance cores.
  \item The end-to-end scripts delete target result directories before writing; back up results if needed.
  \item If you encounter an \texttt{awk} error in \texttt{bench-gemm.sh}, replace \texttt{test-to-csv.sh} with \texttt{test-to-csv-backup.sh} in the script.
\end{itemize}


\end{document}